\documentclass[conference]{IEEEtran}
\IEEEoverridecommandlockouts

\usepackage[T1]{fontenc}
\usepackage{graphicx}
\usepackage{booktabs}
\usepackage{array}
\usepackage{amsmath}
\usepackage{amssymb}
\usepackage{xurl}
\usepackage[colorlinks=true, allcolors=blue]{hyperref}
\usepackage{cite}
\usepackage{orcidlink}

\graphicspath{{figs/}}
\newcommand{\dgxARPrefillOneKSavingMs}{207}
\newcommand{\dgxARPrefillShortCostMs}{41}
\newcommand{\dgxARStepCostRdmaMs}{1.7}

\newcommand{\dgxARStepCostTcpMs}{9.1}
\newcommand{\dgxARStepSavingMs}{7.4}

\newcommand{\dgxAuditDevicesWord}{four}

\newcommand{\dgxAuditIBChannels}{25{,}040}
\newcommand{\dgxAuditLaunches}{148}

\newcommand{\dgxAuditVerified}{148}

\newcommand{\dgxBandARRdmaOverTcpMax}{9.0}
\newcommand{\dgxBandARRdmaOverTcpMin}{2.4}

\newcommand{\dgxBoundDenseDirectMed}{1.25}
\newcommand{\dgxBoundDensePPMax}{2.54}
\newcommand{\dgxBoundDensePPMed}{2.40}
\newcommand{\dgxBoundDensePPMin}{2.08}
\newcommand{\dgxBoundDenseSingleMax}{1.27}
\newcommand{\dgxBoundDenseSingleMin}{1.12}
\newcommand{\dgxBoundDenseTPMax}{1.41}
\newcommand{\dgxBoundDenseTPMin}{1.24}

\newcommand{\dgxBoundMoeSingleMax}{0.52}
\newcommand{\dgxBoundMoeSingleMin}{0.11}

\newcommand{\dgxBudgetLogs}{208}

\newcommand{\dgxCVEnergyMedian}{0.5}

\newcommand{\dgxCVITLMedian}{0.4}

\newcommand{\dgxCVOddLaunchDevMax}{19}
\newcommand{\dgxCVOddLaunchDevMin}{9}
\newcommand{\dgxCVPairSpreadMax}{0.4}

\newcommand{\dgxCVStrictMax}{86.6}

\newcommand{\dgxCVTTFTMedian}{0.5}

\newcommand{\dgxCVThrCountOverFiveWord}{seven}

\newcommand{\dgxCVThrMedian}{0.4}
\newcommand{\dgxCVThrPNinety}{2.1}
\newcommand{\dgxCampaignEnd}{2026-11-10}
\newcommand{\dgxCampaignStart}{2026-10-31}
\newcommand{\dgxCampaignWallHours}{251}

\newcommand{\dgxCellsEnergyPPLowerThanTPWord}{four}
\newcommand{\dgxCellsEnergySingleLowerThanPP}{78}
\newcommand{\dgxCellsEnergySingleLowerThanTP}{78}
\newcommand{\dgxCellsEnergyTPLowerThanPP}{89}

\newcommand{\dgxCellsEnergyTPPPUnsepWord}{three}

\newcommand{\dgxCellsITLTPPPUnsepWord}{six}
\newcommand{\dgxCellsITLTPlower}{90}
\newcommand{\dgxCellsSingle}{78}
\newcommand{\dgxCellsTPPP}{96}

\newcommand{\dgxCellsThrPPgtSingle}{60}

\newcommand{\dgxCellsThrSinglegtPPMaxLoad}{4}
\newcommand{\dgxCellsThrSinglegtPPWord}{six}

\newcommand{\dgxCellsThrTPPPUnsepLoads}{64}
\newcommand{\dgxCellsThrTPPPUnsepWord}{three}

\newcommand{\dgxCellsThrTPgtPP}{93}
\newcommand{\dgxCellsThrTPgtSingle}{78}
\newcommand{\dgxCurvesPeakAtTop}{43}
\newcommand{\dgxDualOnlyPPTPITLRatioMax}{1.97}

\newcommand{\dgxDualOnlyPPTPITLRatioMin}{1.52}
\newcommand{\dgxDualOnlyPPTPRatioMax}{0.86}

\newcommand{\dgxDualOnlyPPTPRatioMin}{0.79}
\newcommand{\dgxEffBWDenseSingleMax}{244}
\newcommand{\dgxEffBWDenseSingleMin}{216}
\newcommand{\dgxEnergyFloorMax}{2.22}
\newcommand{\dgxEnergyFloorMaxModel}{Mistral-Medium-3.5-128B NVFP4}
\newcommand{\dgxEnergyFloorMin}{0.17}
\newcommand{\dgxEnergyFloorMinModel}{Gemma-4-26B-A4B NVFP4}
\newcommand{\dgxEnergyFloorSpread}{13}
\newcommand{\dgxEnergyMinPPBelowTPCountWord}{three}

\newcommand{\dgxEnergyPPAtOneMed}{1.51}

\newcommand{\dgxEnergyPPAtSixtyFourMax}{1.33}
\newcommand{\dgxEnergyPPAtSixtyFourMed}{1.19}
\newcommand{\dgxEnergyPPAtSixtyFourMin}{1.05}

\newcommand{\dgxEnergyRiseCurvesWord}{six}

\newcommand{\dgxEnergyTPAtOneMed}{1.15}

\newcommand{\dgxEnergyTPAtSixtyFourMax}{1.31}
\newcommand{\dgxEnergyTPAtSixtyFourMed}{1.18}
\newcommand{\dgxEnergyTPAtSixtyFourMin}{1.09}

\newcommand{\dgxFallingSingleCountWord}{three}
\newcommand{\dgxFallingSingleNames}{Llama-3.1-8B BF16, Qwen3.6-27B FP8, and Qwen3.6-35B-A3B NVFP4}

\newcommand{\dgxFallingTPCountWord}{one}
\newcommand{\dgxFallingTPNames}{Llama-3.1-8B BF16}
\newcommand{\dgxGPUUtil}{0.80}

\newcommand{\dgxGemmaDenseSinglePeak}{93.7}

\newcommand{\dgxGemmaDenseTPITLOne}{84}

\newcommand{\dgxGemmaMoeSinglePeak}{381}

\newcommand{\dgxGptOssPPTPRatio}{0.74}

\newcommand{\dgxITLOneMax}{229}
\newcommand{\dgxITLOneMaxModel}{Qwen3.6-27B BF16}
\newcommand{\dgxITLOneMin}{14}
\newcommand{\dgxITLOneMinModel}{Qwen3.6-35B-A3B NVFP4}

\newcommand{\dgxKVUsageMaxConfig}{on one Spark}
\newcommand{\dgxKVUsageMaxLoad}{64}
\newcommand{\dgxKVUsageMaxModel}{Qwen3.6-27B BF16}
\newcommand{\dgxKVUsageMaxPct}{68}
\newcommand{\dgxLadderBFTPITLRatioMax}{0.60}
\newcommand{\dgxLadderBFTPITLRatioMin}{0.52}

\newcommand{\dgxLadderITLRatioMax}{0.40}
\newcommand{\dgxLadderITLRatioMin}{0.35}

\newcommand{\dgxLadderLlamaPeakGain}{2.97}
\newcommand{\dgxLadderPeakGainMax}{2.97}
\newcommand{\dgxLadderPeakGainMin}{1.26}

\newcommand{\dgxLadderQwenDensePeakGain}{1.26}

\newcommand{\dgxLadderQwenMoePeakGain}{1.59}

\newcommand{\dgxLlamaAllReducesPerStep}{64}

\newcommand{\dgxLlamaBFPPITLSixtyFour}{311}

\newcommand{\dgxLlamaBFPPThrOne}{13.1}

\newcommand{\dgxLlamaBFSingleBucketTTFTShort}{86}

\newcommand{\dgxLlamaBFSingleITLSixtyFour}{521}

\newcommand{\dgxLlamaBFSingleThrOne}{13.2}

\newcommand{\dgxLlamaBFTPITLRatio}{0.52}

\newcommand{\dgxLlamaBFTPRatio}{1.66}

\newcommand{\dgxLlamaBFTPThrOne}{25.0}

\newcommand{\dgxLlamaFPTPITLRatio}{0.55}

\newcommand{\dgxLlamaNVSingleBucketTTFTShort}{37}

\newcommand{\dgxLlamaNVTPBucketTTFTShort}{83}

\newcommand{\dgxLlamaNVTPITLRatio}{0.58}

\newcommand{\dgxLlamaNVTPRatio}{1.36}

\newcommand{\dgxLlamaTPShortAddMax}{47}
\newcommand{\dgxLlamaTPShortAddMin}{24}
\newcommand{\dgxLongBucketCurvesUnderTen}{26}
\newcommand{\dgxLongBucketMinRequests}{3}
\newcommand{\dgxLooseSLOms}{200}

\newcommand{\dgxMarlinNVCountWord}{three}
\newcommand{\dgxMarlinNVNames}{Qwen3.6-27B, Qwen3.6-35B-A3B, and Nemotron-3.5-Lightning-30B-A3B}
\newcommand{\dgxMaxEndToEnd}{717}
\newcommand{\dgxMaxEndToEndConfig}{PP=2}
\newcommand{\dgxMaxEndToEndLoad}{64}
\newcommand{\dgxMaxEndToEndModel}{Mistral-Medium-3.5-128B NVFP4}
\newcommand{\dgxMaxModelLen}{8{,}704}
\newcommand{\dgxMaxWarmupS}{846}
\newcommand{\dgxMaxWindowS}{2{,}280}
\newcommand{\dgxMeasureHours}{145}
\newcommand{\dgxMinWindowRequests}{20}

\newcommand{\dgxMistralPPITLOne}{423}

\newcommand{\dgxMistralPPPeak}{51.5}

\newcommand{\dgxMistralTPITLOne}{215}

\newcommand{\dgxMistralTPPeak}{65.1}

\newcommand{\dgxMistralTotalGiB}{84}
\newcommand{\dgxMixInOutRatio}{10}
\newcommand{\dgxMixMeanInput}{2{,}670}
\newcommand{\dgxMixMeanOutput}{256}
\newcommand{\dgxMixMeanTotal}{2{,}926}

\newcommand{\dgxNativeNVCountWord}{five}

\newcommand{\dgxNumCampaignWindows}{918}
\newcommand{\dgxNumCheckpoints}{17}
\newcommand{\dgxNumComplete}{13}

\newcommand{\dgxNumCompleteWord}{thirteen}
\newcommand{\dgxNumCurves}{47}

\newcommand{\dgxNumDualCurves}{33}

\newcommand{\dgxNumDualOnlyWord}{three}
\newcommand{\dgxNumExcludedWindows}{72}
\newcommand{\dgxNumExtendedWindows}{30}

\newcommand{\dgxNumLaunches}{141}

\newcommand{\dgxNumLoadsWord}{six}

\newcommand{\dgxNumPoints}{282}

\newcommand{\dgxNumReplicatesWord}{three}
\newcommand{\dgxNumRequests}{272{,}630}

\newcommand{\dgxNumWindows}{846}

\newcommand{\dgxPPBeatsTPEightKCountWord}{ten}

\newcommand{\dgxPPBucketRatioEightKMax}{0.69}
\newcommand{\dgxPPBucketRatioEightKMin}{0.63}
\newcommand{\dgxPPBucketRatioFourKMax}{0.84}
\newcommand{\dgxPPBucketRatioFourKMin}{0.76}
\newcommand{\dgxPPBucketRatioOneKMax}{1.10}
\newcommand{\dgxPPBucketRatioOneKMin}{1.00}
\newcommand{\dgxPPBucketRatioShortMax}{1.01}
\newcommand{\dgxPPBucketRatioShortMin}{0.95}

\newcommand{\dgxPPITLPenaltyMaxMs}{3.8}
\newcommand{\dgxPPITLRatioMax}{1.03}

\newcommand{\dgxPPITLRatioMin}{0.99}
\newcommand{\dgxPPKVRatioMax}{4.11}

\newcommand{\dgxPPKVRatioMin}{1.89}
\newcommand{\dgxPPPrefillRatioMax}{1.68}
\newcommand{\dgxPPPrefillRatioMed}{1.62}
\newcommand{\dgxPPPrefillRatioMin}{1.48}
\newcommand{\dgxPPRatioMax}{1.65}
\newcommand{\dgxPPRatioMed}{1.39}
\newcommand{\dgxPPRatioMin}{1.17}
\newcommand{\dgxPPStrictRatioMax}{1.77}
\newcommand{\dgxPPStrictRatioMed}{1.23}
\newcommand{\dgxPPStrictRatioMin}{1.08}

\newcommand{\dgxPPTPRatioMax}{1.02}
\newcommand{\dgxPPTPRatioMed}{0.93}
\newcommand{\dgxPPTPRatioMin}{0.74}
\newcommand{\dgxPPTTFTAtSixtyFourRatioMed}{0.66}
\newcommand{\dgxPPTTFTRatioMax}{1.10}

\newcommand{\dgxPPTTFTRatioMin}{0.99}

\newcommand{\dgxPowerGapHeadMax}{0.18}
\newcommand{\dgxPowerGapWorkerMax}{1.04}

\newcommand{\dgxPowerRateHead}{10}
\newcommand{\dgxPowerRateWorker}{1.0}

\newcommand{\dgxPrefillRsqMin}{0.986}

\newcommand{\dgxQwenDenseBFSinglePrefill}{1{,}248}

\newcommand{\dgxQwenDenseBFTPITLOne}{123}

\newcommand{\dgxQwenDenseBFTPRatio}{1.65}

\newcommand{\dgxQwenDenseFPSingleITLOne}{130}

\newcommand{\dgxQwenDenseFPSinglePeak}{81.6}

\newcommand{\dgxQwenDenseFPSinglePrefill}{1{,}975}

\newcommand{\dgxQwenDenseFPTPITLOne}{70}

\newcommand{\dgxQwenDenseNVSingleITLOne}{84}

\newcommand{\dgxQwenDenseNVSinglePeak}{81.1}

\newcommand{\dgxQwenDenseNVSinglePrefill}{1{,}129}

\newcommand{\dgxQwenDenseNVTPRatio}{1.63}

\newcommand{\dgxQwenMoeBFTPRatio}{1.65}

\newcommand{\dgxQwenMoeNVPPTPRatio}{1.02}

\newcommand{\dgxQwenMoeNVTPRatio}{1.51}

\newcommand{\dgxQwenPairITLRatioMax}{6.8}
\newcommand{\dgxQwenPairITLRatioMin}{6.2}
\newcommand{\dgxQwenPairPeakRatioMax}{3.0}
\newcommand{\dgxQwenPairPeakRatioMin}{2.4}

\newcommand{\dgxRdmaARLatEightKiBMin}{27}

\newcommand{\dgxRdmaARPlateauMaxUs}{658}
\newcommand{\dgxRdmaARPlateauMinUs}{593}

\newcommand{\dgxRepEnergyTPHigherSixtyFour}{12}

\newcommand{\dgxRepEnergyTPLowerEight}{10}

\newcommand{\dgxRepITLTPHigherSixtyFour}{12}

\newcommand{\dgxRepITLTPLowerEight}{10}

\newcommand{\dgxRepLoadMin}{8}
\newcommand{\dgxRepThrPPRatioMax}{0.92}
\newcommand{\dgxRepThrTPAboveEight}{10}

\newcommand{\dgxRepThrTPAboveSixtyFourNames}{gpt-oss-120b MXFP4}

\newcommand{\dgxRepThrTPBelowSixtyFour}{11}
\newcommand{\dgxRepThrTPBelowSixtyFourWord}{eleven}
\newcommand{\dgxRepThrTPRatioEightMax}{1.55}
\newcommand{\dgxRepThrTPRatioEightMin}{0.95}
\newcommand{\dgxRepThrTPRatioSixtyFourMax}{1.05}
\newcommand{\dgxRepThrTPRatioSixtyFourMin}{0.75}

\newcommand{\dgxRepThrTPUndecidedSixtyFourNames}{Gemma-4-31B NVFP4}
\newcommand{\dgxSealedArtifacts}{3{,}952}
\newcommand{\dgxSealedChecked}{972}

\newcommand{\dgxSinglePowerGrowthMax}{1.8}
\newcommand{\dgxSinglePowerGrowthMin}{1.2}
\newcommand{\dgxSingleThrGrowthMax}{15}
\newcommand{\dgxSingleThrGrowthMin}{3.7}
\newcommand{\dgxSpecBW}{273}
\newcommand{\dgxStrictCliffCurves}{38}
\newcommand{\dgxStrictCliffITLMax}{173}
\newcommand{\dgxStrictCliffITLMin}{107}

\newcommand{\dgxStrictCliffTTFTCountWord}{two}
\newcommand{\dgxStrictDualOnlySepNames}{Qwen3-235B-A22B NVFP4}
\newcommand{\dgxStrictDualOnlyUndecidedNames}{Qwen3-Next-80B-A3B BF16}

\newcommand{\dgxStrictNonzeroCount}{12}
\newcommand{\dgxStrictNonzeroCountWord}{twelve}

\newcommand{\dgxStrictNonzeroSingleCountWord}{ten}

\newcommand{\dgxStrictOnlyTPCountWord}{two}
\newcommand{\dgxStrictOnlyTPNames}{Qwen3.6-27B FP8 and Gemma-4-31B NVFP4}
\newcommand{\dgxStrictRangeTPgtPPCount}{7}
\newcommand{\dgxStrictRangeTPgtSingleCount}{11}

\newcommand{\dgxStrictRepAboveCountWord}{four}
\newcommand{\dgxStrictRepAboveNames}{Llama-3.1-8B BF16, Qwen3.6-27B NVFP4, Qwen3.6-35B-A3B BF16, and gpt-oss-120b MXFP4}
\newcommand{\dgxStrictRepAheadAllCountWord}{six}

\newcommand{\dgxStrictRepBelowCountWord}{four}

\newcommand{\dgxStrictRepUndecidedCountWord}{two}
\newcommand{\dgxStrictRepUndecidedNames}{Llama-3.1-8B FP8 and Nemotron-3.5-Lightning-30B-A3B NVFP4}
\newcommand{\dgxStrictSLOms}{100}
\newcommand{\dgxStrictSepOwnBestCount}{12}
\newcommand{\dgxStrictSepSameLoadCount}{10}

\newcommand{\dgxStrictUnsepSameLoadNames}{Llama-3.1-8B FP8 and Nemotron-3.5-Lightning-30B-A3B NVFP4}

\newcommand{\dgxTPBestLooseCount}{12}
\newcommand{\dgxTPBestPeakCount}{12}
\newcommand{\dgxTPBestStrictCount}{12}
\newcommand{\dgxTPBucketRatioEightKMax}{0.82}
\newcommand{\dgxTPBucketRatioEightKMin}{0.60}

\newcommand{\dgxTPITLRatioDenseMax}{0.58}
\newcommand{\dgxTPITLRatioDenseMin}{0.52}
\newcommand{\dgxTPITLRatioMax}{0.73}
\newcommand{\dgxTPITLRatioMed}{0.58}
\newcommand{\dgxTPITLRatioMin}{0.52}
\newcommand{\dgxTPITLRatioMoeMax}{0.73}
\newcommand{\dgxTPITLRatioMoeMin}{0.60}
\newcommand{\dgxTPKVRatioMax}{4.38}

\newcommand{\dgxTPKVRatioMin}{2.09}
\newcommand{\dgxTPLowerITLCount}{13}
\newcommand{\dgxTPPrefillRatioMax}{1.70}
\newcommand{\dgxTPPrefillRatioMed}{1.49}
\newcommand{\dgxTPPrefillRatioMin}{1.29}
\newcommand{\dgxTPRatioMax}{1.68}
\newcommand{\dgxTPRatioMed}{1.59}
\newcommand{\dgxTPRatioMin}{1.36}

\newcommand{\dgxTPStrictRatioMax}{6.62}
\newcommand{\dgxTPStrictRatioMed}{2.00}
\newcommand{\dgxTPStrictRatioMin}{1.39}
\newcommand{\dgxTPTTFTAtSixtyFourRatioMed}{0.65}
\newcommand{\dgxTPTTFTRatioMax}{0.99}

\newcommand{\dgxTPTTFTRatioMin}{0.60}
\newcommand{\dgxTTFTNonMonotoneCurves}{12}

\newcommand{\dgxTTFTSLOs}{3}

\newcommand{\dgxTcpARLatEightKiBMin}{143}

\newcommand{\dgxTempHeadMax}{88}
\newcommand{\dgxTempWorkerMax}{87}
\newcommand{\dgxVLLMVersion}{0.29.0}

\newcommand{\dgxWeightOnlyCountWord}{four}
\newcommand{\dgxWeightOnlyNativeNames}{Mistral-Medium-3.5-128B NVFP4}
\newcommand{\dgxWindowS}{600}

\newcommand{\dgxXLlamaBFPPPeakRatio}{1.04}

\newcommand{\dgxXLlamaBFTPITLDropMs}{11}

\newcommand{\dgxXLlamaBFTPPeakRatio}{1.68}

\newcommand{\dgxXLlamaFPPPPeakRatio}{1.07}

\newcommand{\dgxXLlamaFPTPITLDropMs}{7}

\newcommand{\dgxXLlamaFPTPPeakRatio}{2.37}

\newcommand{\dgxXLlamaNVTPITLDropMs}{10}

\newcommand{\dgxXLlamaNVTPPeakRatio}{2.58}

\newcommand{\dgxXLlamaTPTTFTDropMsMax}{246}
\newcommand{\dgxXLlamaTPTTFTDropMsMin}{231}
\newcommand{\dgxXLlamaTPTTFTRatioMax}{0.40}
\newcommand{\dgxXLlamaTPTTFTRatioMin}{0.28}

\newcommand{\tp}{TP$=$2}
\newcommand{\pp}{PP$=$2}
\newcommand{\x}{$\times$}

\newcounter{finding}
\newcommand{\finding}[1]{%
  \refstepcounter{finding}%
  \par\vspace{4pt}\noindent
  \fbox{\parbox{\dimexpr\linewidth-2\fboxsep-2\fboxrule\relax}{\textbf{Finding \thefinding.} #1}}%
  \par\vspace{4pt}}

\begin{document}

\title{Benchmarking LLM Inference on Desktop-Scale NVIDIA Grace Blackwell Clusters}

\author{
\IEEEauthorblockN{
Kevin Bhimani\,\orcidlink{0000-0003-3695-3164}\IEEEauthorrefmark{1}\IEEEauthorrefmark{2},
Christian Escobar\,\orcidlink{0000-0003-2907-7311}\IEEEauthorrefmark{1},
Daniel Mihovch\,\orcidlink{0009-0005-3395-258X}\IEEEauthorrefmark{1}, 
Rudolf Eigenmann\,\orcidlink{0000-0003-1651-827X}
and Sunita Chandrasekaran\,\orcidlink{0000-0002-3560-9428}
}
\IEEEauthorblockA{
University of Delaware, Newark, Delaware, USA\\
\IEEEauthorrefmark{1}These authors contributed equally.\quad
\IEEEauthorrefmark{2}Corresponding author: kbhimani@udel.edu
}
}

\maketitle
\pagestyle{plain}
\thispagestyle{plain}


\begin{abstract}
Two NVIDIA DGX Spark GB10 systems connect directly over 200 GbE to serve larger language models and increase inference capacity, but the benefits depend on how they are deployed. We benchmark \dgxNumCheckpoints\ open-weight checkpoints spanning dense and mixture-of-experts architectures, 8B--235B parameters and four weight precisions on one and two RDMA-connected systems. Using vLLM, we compare tensor parallelism (\tp), pipeline parallelism (\pp) and a two-replica baseline at concurrency 1--64, with three independent launches per configuration. \tp\ delivers a median \dgxTPRatioMed\x\ peak-throughput gain over one Spark, versus \dgxPPRatioMed\x\ for \pp\ and has higher throughput in \dgxCellsThrTPgtPP\ of \dgxCellsTPPP\ matched comparisons with separated 95\% Welch intervals. It reduces unloaded inter-token latency to \dgxTPITLRatioMin--\dgxTPITLRatioMax\x\ of one Spark, whereas \pp\ provides little reduction. \tp\ also achieves the highest goodput under a strict \dgxStrictSLOms\,ms inter-token latency objective wherever both strategies are supported and the objective is met. For models fitting on one Spark, \tp\ generally favors low-concurrency serving, while replication offers higher throughput at high concurrency. One Spark uses less GPU energy per output token at matched concurrency. Across three precision ladders, NVFP4 reduces unloaded latency more than adding a second Spark at BF16. NCCL microbenchmarks and comparisons with an earlier TCP campaign suggest that collective communication overhead contributes to the advantage of \tp. These results provide deployment guidance for desktop-scale inference.
\end{abstract}


\section{Introduction}
\label{sec:intro}
Compact local GPU systems are becoming an attractive platform for organizations seeking to serve large language models (LLMs) on premises, particularly when data privacy, infrastructure costs and control over deployment are important considerations. The NVIDIA DGX Spark integrates a GB10 system-on-chip, comprising a Blackwell GPU and a 20-core Arm CPU sharing 128\,GB of LPDDR5x memory, into a desktop-scale system. Systems of this size enable local inference with open-weight language models without requiring datacenter-scale hardware. Two DGX Sparks can be connected directly through their ConnectX-7 network ports, doubling the memory capacity. This enables the deployment of models that cannot fit on a single system and provides additional resources for serving concurrent requests. However, the performance benefits depend on how those resources are used. This paper investigates the tradeoffs between distributing a model across two Sparks and deploying independent replicas, evaluating throughput, latency, goodput and GPU energy efficiency.

A model can be distributed across two GPUs using two standard approaches. Tensor parallelism (\tp) partitions the computation within each layer across both GPUs and requires collective communication during every decode step. Pipeline parallelism (\pp) assigns different layers to each GPU and transfers activations between pipeline stages. Conventional multi-GPU serving guidance favors tensor parallelism within a server, where high-bandwidth interconnects such as NVLink support frequent collective operations and pipeline parallelism across servers to minimize communication overhead \cite{narayanan2021}. DGX Spark presents a different hardware configuration. The two GPUs communicate through a pair of 200\,GbE connections rather than NVLink, the platform does not support GPUDirect RDMA \cite{nvidia2026sparkporting} and each GPU accesses unified LPDDR5x memory instead of dedicated high-bandwidth memory (HBM). These characteristics make the relative benefits of tensor and pipeline parallelism difficult to predict from conventional datacenter experience. We evaluate both strategies using NCCL over RDMA over Converged Ethernet version 2 (RoCEv2), with all four network rails configured according to NVIDIA's recommended setup \cite{nvidia2026sparkplaybook}. We also use an earlier TCP-based campaign as supporting evidence to examine the role of inter-node communication overhead.

\begin{figure}[!tp]
\centering
\includegraphics[width=\columnwidth]{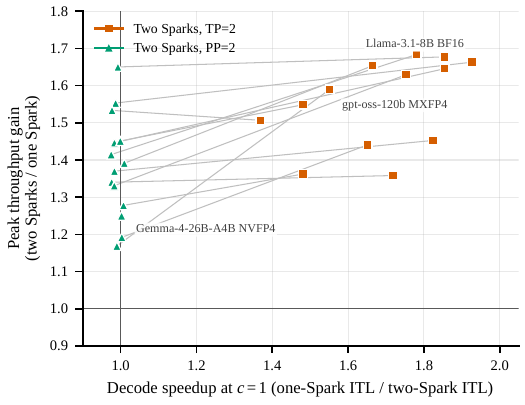}
\caption{Performance gains from distributing a model across two DGX Sparks. Each marker represents one checkpoint under one parallel deployment, with lines connecting \tp\ and \pp\ results for the same checkpoint. The horizontal axis shows unloaded decode speedup, defined as the ratio of one-Spark to two-Spark inter-token latency at concurrency 1. The vertical axis shows peak throughput speedup, defined as the ratio of two-Spark to one-Spark peak throughput over concurrency levels up to 64. \tp\ improves both decode latency and peak throughput, whereas \pp\ improves peak throughput with little change in unloaded decode latency.}
\label{fig:gainmap}
\end{figure}

The study covers \dgxNumCheckpoints\ open-weight checkpoints spanning dense and mixture-of-experts (MoE) architectures, four weight precisions and 8B to 235B parameters. Of these, \dgxNumDualOnlyWord\ require two Sparks for the evaluated configurations. We evaluate up to three deployments per checkpoint: one Spark, two Sparks with \tp\ and two Sparks with \pp. Each deployment is measured at \dgxNumLoadsWord\ concurrency levels, with \dgxNumReplicatesWord\ independent server launches per measurement point. The resulting campaign comprises \dgxNumWindows\ measurement windows and \dgxMeasureHours\ measurement hours across \dgxNumLaunches\ launches, without a failed request in the analyzed windows. For models that fit on one Spark, we additionally derive the performance of two independent replicas from single-Spark measurements to determine when splitting a model is preferable to replication. Fig.~\ref{fig:gainmap} previews the performance gains from distributing a model across two Sparks.

The paper makes four contributions.
\begin{itemize}

\item \textbf{A reproducible benchmark of LLM inference on DGX Spark}
(Sections~\ref{sec:method}--\ref{sec:results}).
We evaluate \dgxNumCheckpoints\ open-weight checkpoints spanning multiple architectures, model sizes and weight precisions across single- and dual-node deployments. Every measurement point is replicated across three independent server launches, NCCL transport is verified from the logs of every dual-node launch and all reported results are generated from sealed measurements (Appendix~\ref{app:artifact}).

\item \textbf{A systematic comparison of tensor and pipeline parallelism over RDMA}
(Sections~\ref{sec:throughput}--\ref{sec:energy}).
We show that \tp\ outperforms \pp\ in peak throughput, unloaded decode latency and goodput under a strict inter-token latency objective. Welch confidence intervals place \tp\ above \pp\ in \dgxCellsThrTPgtPP\ of \dgxCellsTPPP\ throughput comparisons, with none favoring \pp. We also identify the conditions under which \pp\ is beneficial, particularly during prefill of prompts spanning multiple chunks.

\item \textbf{An evaluation of model splitting versus independent replication}
(Section~\ref{sec:replicas}).
For models that fit on one Spark, we compare distributed inference with two independent replicas derived from single-node measurements. We find that \tp\ provides higher throughput at low concurrency, while replication generally performs better at high concurrency. \pp\ does not significantly outperform replication and a single Spark consistently provides the lowest GPU energy per generated token at matched concurrency. These results inform deployment choices based on workload and performance objectives.

\item \textbf{An analysis of the hardware and communication factors governing inference performance}
(Sections~\ref{sec:levers} and \ref{sec:transport}).
We relate unloaded decode latency to memory bandwidth, weight precision, model architecture and inter-node communication. Dense-model decoding operates at \dgxBoundDenseSingleMin--\dgxBoundDenseSingleMax\x\ of the theoretical weight-read bound and reducing precision from BF16 to NVFP4 improves unloaded latency more than adding a second Spark under either parallel strategy. NCCL microbenchmarks and comparisons with an earlier TCP-based campaign provide supporting evidence that lower collective communication overhead contributes to the observed advantage of \tp\ over \pp.

\end{itemize}

\section{Background}
\label{sec:background}
\subsection{The platform}
The GB10 system-on-chip integrates a 20-core Arm CPU and a Blackwell GPU sharing 128\,GB of unified LPDDR5x memory with a nominal bandwidth of \dgxSpecBW\,GB/s \cite{nvidia2025sparkhw}. Unlike datacenter GPUs with dedicated high-bandwidth memory (HBM), the GB10 GPU accesses shared system memory. At small batch sizes, dense-model decoding is primarily memory-bandwidth-bound, with a theoretical lower bound on decode step time given by the model's weight footprint divided by available memory bandwidth \cite{williams2009roofline,pope2023efficiently,yuan2024llminferenceunveiledsurvey}.

Two DGX Sparks can connect directly through their ConnectX-7 ports without a network switch \cite{nvidia2025sparkstack,nvidia2026sparkplaybook}. Each Ethernet interface has a corresponding RDMA interface \cite{nvidia2025sparkstack}, allowing NVIDIA's Collective Communications Library (NCCL) \cite{nvidia2026nccl} to communicate over RDMA over Converged Ethernet version 2 (RoCEv2) \cite{ibta2014rocev2} instead of TCP sockets. When RDMA devices are unavailable inside the container, NCCL can silently fall back to TCP. We therefore verify the transport used by every dual-node launch (Section~\ref{sec:setup}).

\subsection{Serving pipeline}
We serve models using vLLM \cite{kwon2023vllm}, which supports continuous batching \cite{yu2022orca} and paged key-value (KV) cache management. Inference consists of two phases: prefill, which processes the input prompt and generates the first token and decode, which generates subsequent tokens autoregressively. Prefill is typically compute-bound for long prompts and largely determines time to first token (TTFT). With chunked prefill \cite{agrawal2023sarathi}, vLLM divides long prompts into chunks based on a token budget and schedules them alongside decoding requests. Decode is typically memory-bandwidth-bound at small batch sizes, as each step accesses model weights and the active KV cache and determines inter-token latency (ITL). Since maximum throughput can occur at loads with unacceptable interactive latency, we also evaluate goodput and GPU energy per generated token. Following the concept introduced by DistServe \cite{zhong2024distserve}, we define goodput at each load as the rate of completed requests that satisfy both TTFT and ITL objectives.

\subsection{Two ways to split one model}

\begin{figure}[!tp]
\centering
\includegraphics[width=\columnwidth]{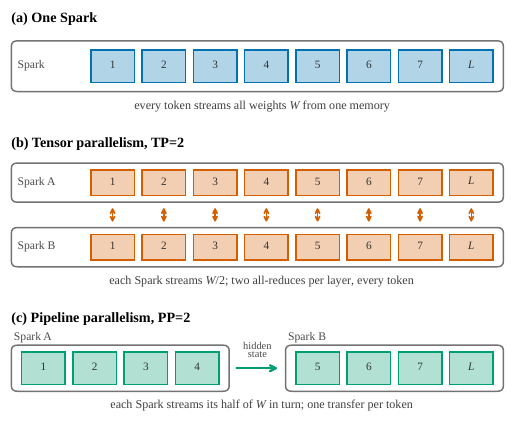}
\caption{Three serving deployments. (a) One Spark executes all layers and reads the model weights $W$ from local memory. (b) \tp\ distributes each layer across both Sparks, requiring two all-reduces per layer. (c) \pp\ assigns half the layers to each Spark and transfers activations between stages.}
\label{fig:deployments}
\end{figure}

Fig.~\ref{fig:deployments} illustrates the three deployments. \tp\ shards the weight matrices of each layer across both GPUs \cite{shoeybi2019megatron,narayanan2021}. Each node reads approximately half the weights per decode step, with an all-reduce after the attention and feed-forward blocks. A dense model with $L$ layers therefore requires approximately $2L$ all-reduces per step. Each collective operates on an activation tensor proportional to the batch size and hidden dimension. For BF16 activations and a hidden size of 4096, as in Llama-3.1-8B, the payload is 8\,KiB per decode token, or 512\,KiB for a batch of 64 tokens. A prefill chunk of 2{,}048 tokens produces a 16\,MiB payload. At low concurrency, collective latency dominates communication overhead, while larger batches increase the importance of message size (Section~\ref{sec:transport}).

\pp\ assigns the first half of the layers to one GPU and the remaining layers to the other \cite{huang2019gpipe}. Each node reads only its stage's weights and transfers intermediate activations across the stage boundary. For Llama-style models in vLLM, the transferred hidden state and residual together require twice the payload of one TP all-reduce. At concurrency 1, the stages execute sequentially, so pipeline parallelism cannot reduce decode latency through stage overlap. With multiple requests in flight or long prompts divided into prefill chunks, the stages can process different batches concurrently, improving throughput. Both strategies distribute KV-cache storage across the two nodes, with \tp\ partitioning attention heads and \pp\ partitioning layers.

\subsection{Precision and sparsity}
Weight quantization reduces memory traffic during inference. FP8 (E4M3) approximately halves the BF16 weight footprint \cite{micikevicius2022fp8formatsdeeplearning}, while NVFP4 uses 4-bit weights with two-level block scaling, further reducing memory requirements and supporting native execution on Blackwell GPUs \cite{alvarez2025nvfp4}. MXFP4 is a separate 4-bit microscaling format used by the gpt-oss mixture-of-experts (MoE) models \cite{openai2025gptoss}. Several NVIDIA ModelOpt checkpoints \cite{nvidia2025modelopt} also use an FP8 key-value (KV) cache (Table~\ref{tab:models}), so comparisons across precision variants may reflect changes in both weight and KV-cache precision. MoE architectures \cite{shazeer2017moe,fedus2022switch} activate only a subset of experts for each token, reducing weight memory traffic relative to dense models with similar total parameter counts.

\section{Methodology}
\label{sec:method}

Table~\ref{tab:metrics} defines the workload variables and the metrics and the reasons for the choices.

\begin{table*}[!tp]
\caption{Workload variables and reported metrics. Every value in the paper is the mean over three replicate launches unless stated otherwise.}
\label{tab:metrics}
\centering
\footnotesize
\setlength{\tabcolsep}{4pt}
\renewcommand{\arraystretch}{1.15}
\begin{tabular}{@{}l l p{0.66\textwidth}@{}}
\toprule
Quantity & Unit & Definition \\
\midrule

Measurement window $W$ & s &
Interval over which throughput, goodput, and energy are measured: at least 600\,s, extended in 60\,s steps until at least 20 requests have started inside it (3{,}600\,s cap). \\

Concurrency $c$ & requests &
Requests the client keeps in flight. A new request is issued the moment one completes (closed loop). \\

Replicate & launch &
One fresh server launch followed by a complete ascending sweep over $c$. Every point has three replicates; the order of the three deployments rotates across replicates. \\

Output throughput & tok/s &
Output tokens received by the client during $W$, from all in-flight requests, divided by the length of $W$. \\

Peak throughput & tok/s &
Largest replicate-mean output throughput of a deployment over $c\in\{1,\ldots,64\}$; on \dgxCurvesPeakAtTop\ of the \dgxNumCurves\ curves it is the value at $c{=}64$. \\

TTFT & s &
Time to first token: request submission to arrival of the first output token at the client. \\

ITL & ms &
Inter-token latency: the mean gap between consecutive output tokens of one request. The paper reports the median over requests started in $W$. \\

Prefill rate & tok/s &
Reciprocal slope of a linear fit of median TTFT against prompt length over the four workload buckets at $c{=}1$. \\

Goodput & req/s &
Requests started in $W$ with TTFT $\le 3$\,s and mean ITL $\le 100$\,ms (strict tier) or $\le 200$\,ms (loose tier), divided by the length of $W$. \\

Energy per token & J/tok &
GPU energy integrated from sampled device power over $W$, summed over both nodes for two-Spark deployments, divided by the output tokens received in $W$. \\

KV-cache capacity & requests &
Number of concurrent 8{,}704-token requests whose KV cache fits in the memory vLLM reserves at launch, as reported by the engine. \\

Scaling ratio & $\times$ &
A two-Spark deployment's value divided by the one-Spark value of the same checkpoint and metric. Above 1 is better for throughput, goodput, and capacity; below 1 is better for latency and energy. \\

Welch interval & as metric &
95\,\% confidence interval for the difference between two deployments' replicate means (Welch's $t$, $n{=}3$ each). A margin is called separated when the interval excludes zero. \\

\bottomrule
\end{tabular}
\end{table*}

\subsection{Workload}
A measurement point consists of one checkpoint, one deployment and one concurrency level $c\in\{1,4,8,16,32,64\}$. We use four (input, output) token buckets: $(128,128)$, $(1024,256)$, $(4096,256)$ and $(8192,512)$, with weights $1$, $1$, $1$ and $0.5$, respectively. Smooth weighted round robin ensures that the deployments receive the same deterministic bucket sequence at a given concurrency. The workload averages \dgxMixMeanInput\ input tokens and \dgxMixMeanOutput\ output tokens per request, yielding approximately \dgxMixInOutRatio\ prefill tokens per generated token. Reported output throughput counts only generated tokens.

Prompts consist of tokenizer-exact synthetic token sequences with unique eight-token prefixes to prevent prefix reuse. Prefix caching is disabled. Generation uses greedy decoding with fixed output lengths and ignores end-of-sequence tokens. A single asynchronous client on the head node maintains exactly $c$ concurrent requests against vLLM's OpenAI-compatible streaming endpoint, submitting a new request whenever one completes. Output tokens are counted from the returned token identifiers.

\subsection{Windows and accounting}
Each measurement consists of three phases. Warmup lasts at least 120\,s and continues until every client connection completes one request. The measurement window $W$ lasts at least \dgxWindowS\,s and extends in 60\,s increments until at least \dgxMinWindowRequests\ requests have started, with a maximum duration of 3{,}600\,s. Cooldown lasts at least 30\,s, maintaining the workload until all requests admitted during $W$ complete.

Throughput counts output tokens received during $W$ from all in-flight requests, while latency and goodput are calculated from requests admitted during $W$ and tracked to completion. GPU power is sampled using \texttt{nvidia-smi} and integrated over $W$, summing both nodes for two-Spark deployments.

Of the \dgxNumWindows\ measurement windows, \dgxNumExtendedWindows\ required extensions beyond \dgxWindowS\,s, all at $c\le4$ on the slowest checkpoints, reaching at most \dgxMaxWindowS\,s. These windows contain only 20 or 21 admitted requests, with as few as \dgxLongBucketMinRequests\ from the 8{,}192-token bucket.

\subsection{Replicates and intervals}
\label{sec:replicates}

Each replicate consists of a fresh server launch followed by a complete ascending concurrency sweep. Every measurement point has \dgxNumReplicatesWord\ replicates, with deployment order rotated across replicates to mitigate temporal drift. We replicate server launches to capture variability in memory profiling, KV-cache capacity, compilation state and cluster initialization. This variability is small: the median throughput coefficient of variation (CV) across replicates is \dgxCVThrMedian\,\% (Section~\ref{sec:variance}).

Figures report the mean across three launches, with bands showing the minimum and maximum replicate values. We consider a difference between deployments \emph{separated} when its 95\,\% Welch confidence interval excludes zero \cite{welch1947}. These intervals are not adjusted for multiple comparisons. Given the small number of replicates, we interpret separation as evidence of a directional difference rather than definitive statistical significance. Conversely, an undecided comparison does not establish equivalent performance.

\subsection{Objectives}

We evaluate goodput under two service-level objectives (SLOs). Both require time to first token (TTFT) of at most \dgxTTFTSLOs\,s, with mean inter-token latency (ITL) limited to \dgxStrictSLOms\,ms for the strict tier or \dgxLooseSLOms\,ms for the loose tier. Goodput is the rate of requests satisfying both objectives. Unless otherwise stated, reported ITL is the median across requests of their individual mean inter-token latencies. The pooled median of individual token gaps is used only for the memory-bandwidth analysis in Section~\ref{sec:levers}.

\section{Experimental Setup}
\label{sec:setup}

\subsection{Hardware and interconnect}
Our cluster consists of two ASUS Ascent GX10 systems based on the NVIDIA DGX Spark GB10 design. The head node hosts the benchmark client and the server for single-node experiments, while the second node serves as the worker. The head node identifies as an ASUSTeK GX10. Both nodes have matching GPU models, drivers, container images and GPU clocks. Two direct cables connect their ConnectX-7 ports without a switch. Each node exposes four RoCEv2 interfaces, two per physical port, forming four independent rails on separate IPv4 subnets with an MTU of 4096.

We configured NCCL 2.31.2 to use all four RDMA rails and enabled debug logging to verify the communication transport. We audited all \dgxAuditLaunches\ dual-node launches, including canary and failed launches and the excluded checkpoints described in Section~\ref{sec:checkpoints}. All \dgxAuditVerified\ used \texttt{NET/IB} across \dgxAuditDevicesWord\ RoCEv2 devices, with no NCCL channels using TCP sockets or GPUDirect RDMA. 


DGX Spark does not support GPUDirect RDMA because GPU-allocated memory cannot be accessed coherently by PCIe devices \cite{nvidia2026sparkporting}. Consistent with this limitation, NCCL reported \texttt{cuMemGdrSupport 0} in every launch and staged transfers through host-registered buffers. Since CPU and GPU share LPDDR5x memory, staging introduces additional memory copies rather than transfers between separate memory pools. Section~\ref{sec:transport} examines communication performance using NCCL microbenchmarks collected on the same network configuration.

\subsection{Software}

We used NVIDIA's NGC vLLM 26.09 container (vLLM \dgxVLLMVersion, PyTorch 2.14, CUDA 13.4 and FlashInfer 0.6.18) with driver 580.178.04 and Ray 2.56.1 for multi-node coordination \cite{moritz2018ray}. All checkpoints used the same engine settings: a maximum model length of \dgxMaxModelLen\ tokens, GPU memory utilization of \dgxGPUUtil, chunked prefill, CUDA graphs, \texttt{torch.compile} and disabled prefix caching.

In vLLM \dgxVLLMVersion, the default batched-token budget exceeds 2{,}048 only when NVML reports at least 70\,GiB of GPU memory. Because NVML does not report a memory total on GB10, the engine used a 2{,}048-token budget. All \dgxBudgetLogs\ archived server logs confirm this value through \texttt{compile\_ranges\_endpoints}, which reflects the configured token budget in this version.

Under \tp, collective operations used NCCL. We disabled FlashInfer kernel autotuning across all deployments to avoid multi-node startup deadlocks associated with differing kernel configurations \cite{vllm2026pr57197,vllm2026issue52291}. Dual-node deployments used \texttt{-{}-tensor-parallel-size 2} or \texttt{-{}-pipeline-parallel-size 2}, with vLLM's default pipeline schedule. Appendix~\ref{app:config} provides the complete configuration.

\subsection{Checkpoints}
\label{sec:checkpoints}
We evaluate \dgxNumCheckpoints\ model checkpoints, including multiple precision variants of the same underlying models. We refer to each distinct set of model weights and quantization settings as a checkpoint. Table~\ref{tab:models} summarizes the \dgxNumCheckpoints\ checkpoints, spanning dense, mixture-of-experts (MoE) and hybrid architectures. Three model families are evaluated at BF16, FP8 and NVFP4: Llama-3.1-8B \cite{grattafiori2024llama3}, Qwen3.6-27B and Qwen3.6-35B-A3B \cite{qwen2026qwen36}. Gemma-4 contributes a dense 31B and an MoE 26B-A4B model at NVFP4 \cite{gemma2026report,google2026gemma4}. Nemotron-3.5-Lightning-30B-A3B uses a hybrid Mamba-MoE architecture \cite{nvidia2026lightning}, while Qwen3-Next-80B-A3B uses hybrid linear attention \cite{qwen2025qwen3next}. The latter is evaluated at NVFP4 on one and two Sparks and at BF16 on two Sparks only. Additional checkpoints include gpt-oss-120b at MXFP4 \cite{openai2025gptoss}, Mistral-Medium-3.5-128B \cite{mistral2026medium35} and Qwen3-235B-A22B \cite{yang2025qwen3,nvidia2026qwen235nvfp4}. The latter two are evaluated at NVFP4 on two Sparks only (Section~\ref{sec:dualonly}). All Hugging Face revisions were pinned and recorded for reproducibility.

One planned configuration could not be evaluated: \tp\ for Gemma-4-26B-A4B NVFP4, because the FlashInfer CUTLASS MoE implementation in this vLLM version lacks the weight-padding support required for tensor parallelism. The resulting matrix contains \dgxNumCurves\ deployment curves: \dgxNumComplete\ checkpoints under all three deployments, referred to as \emph{complete} checkpoints, Gemma-4-26B-A4B under one Spark and \pp, and \dgxNumDualOnlyWord\ checkpoints under both two-Spark deployments.

Two additional Nemotron-3-Super-120B-A12B checkpoints were measured but excluded from the analysis. NVIDIA's container release notes limit the NVFP4 variant to four sequences on DGX Spark \cite{nvidia2025nemotron3,nvidia2026nemotron3super,nvidia2026vllmrelnotes}. Consequently, measurements above $c{=}4$ primarily reflect scheduler queuing rather than deployment performance. The FP8 variant requires two Sparks, but all three \pp\ launches failed before measurement, leaving no comparable \pp\ results. The \dgxNumExcludedWindows\ excluded windows remain available in the released index of \dgxNumCampaignWindows\ windows.

\begin{table*}[!tp]
\caption{Model checkpoints and deployment configurations, grouped by dense and mixture-of-experts (MoE) architectures. Type indicates active parameters per token for MoE models. Wts and KV denote weight and KV-cache precision. GiB reports the total resident weight footprint across participating nodes. Runs indicates the evaluated deployments: one Spark, \tp\ and \pp. KV capacity is the number of concurrent \dgxMaxModelLen-token sequences supported by vLLM's allocated KV cache. The workload averages \dgxMixMeanTotal\ tokens per request, allowing greater effective concurrency. $^\dagger$ indicates NVFP4 weights served using vLLM's weight-only Marlin kernel (Section~\ref{sec:levers}).}
\label{tab:models}
\centering
\footnotesize
\setlength{\tabcolsep}{3pt}
\begin{tabular}{@{}l l l l l r c r r r@{}}
\toprule
Checkpoint & Type & Attention & Wts & KV & GiB & Runs & \multicolumn{3}{c}{KV capacity (req.)} \\
\cmidrule(lr){8-10}
 & & & & & & & $1\times$ & TP=2 & PP=2 \\
\midrule
Llama-3.1-8B BF16 & dense & full & BF16 & BF16 & 15.0 & \checkmark\checkmark\checkmark & 72 & 163 & 161 \\
Llama-3.1-8B FP8 & dense & full & FP8 & FP8 & 8.5 & \checkmark\checkmark\checkmark & 158 & 337 & 337 \\
Llama-3.1-8B NVFP4 & dense & full & NVFP4 & FP8 & 5.7 & \checkmark\checkmark\checkmark & 164 & 343 & 344 \\
Qwen3.6-27B BF16 & dense & linear hybrid & BF16 & BF16 & 50.2 & \checkmark\checkmark\checkmark & 56 & 187 & 181 \\
Qwen3.6-27B FP8 & dense & linear hybrid & FP8 & BF16 & 27.6 & \checkmark\checkmark\checkmark & 85 & 218 & 212 \\
Qwen3.6-27B NVFP4 & dense & linear hybrid & NVFP4$^{\dagger}$ & FP8 & 19.1 & \checkmark\checkmark\checkmark & 161 & 381 & 368 \\
Gemma-4-31B NVFP4 & dense & sliding window & NVFP4 & FP8 & 30.1 & \checkmark\checkmark\checkmark & 25 & 67 & 48 \\
Mistral-Medium-3.5-128B NVFP4 & dense & full & NVFP4 & FP8 & 84.1 & \textendash\checkmark\checkmark & -- & 66 & 63 \\
Qwen3.6-35B-A3B BF16 & MoE 3B act. & linear hybrid & BF16 & BF16 & 64.7 & \checkmark\checkmark\checkmark & 120 & 492 & 494 \\
Qwen3.6-35B-A3B FP8 & MoE 3B act. & linear hybrid & FP8 & BF16 & 33.6 & \checkmark\checkmark\checkmark & 244 & 621 & 627 \\
Qwen3.6-35B-A3B NVFP4 & MoE 3B act. & linear hybrid & NVFP4$^{\dagger}$ & BF16 & 19.6 & \checkmark\checkmark\checkmark & 301 & 683 & 685 \\
Gemma-4-26B-A4B NVFP4 & MoE 4B act. & sliding window & NVFP4 & FP8 & 17.1 & \checkmark\textendash\checkmark & 129 & -- & 179 \\
Nemotron-3.5-Lightning-30B-A3B NVFP4 & MoE 3B act. & Mamba hybrid & NVFP4$^{\dagger}$ & FP8 & 17.9 & \checkmark\checkmark\checkmark & 899 & 2004 & 2002 \\
Qwen3-Next-80B-A3B NVFP4 & MoE 3B act. & linear hybrid & NVFP4 & FP8 & 44.2 & \checkmark\checkmark\checkmark & 320 & 944 & 951 \\
Qwen3-Next-80B-A3B BF16 & MoE 3B act. & linear hybrid & BF16 & BF16 & 148.6 & \textendash\checkmark\checkmark & -- & 178 & 169 \\
gpt-oss-120b MXFP4 & MoE 5.1B act. & sliding window & MXFP4 & BF16 & 66.1 & \checkmark\checkmark\checkmark & 59 & 258 & 225 \\
Qwen3-235B-A22B NVFP4 & MoE 22B act. & full & NVFP4 & FP8 & 130.0 & \textendash\checkmark\checkmark & -- & 64 & 68 \\
\bottomrule
\end{tabular}

\end{table*}

\subsection{Campaign}
The campaign ran unattended from \dgxCampaignStart\ to \dgxCampaignEnd, spanning \dgxCampaignWallHours\ hours from the first to the last measurement window. Checkpoints were downloaded, benchmarked and removed sequentially. The analyzed windows contain \dgxNumRequests\ completed requests, with no request failures or scheduler preemptions. GPU power measurements were collected from both nodes for every two-Spark window. Following the campaign, a SHA-256 manifest sealed \dgxSealedArtifacts\ artifacts. All \dgxSealedChecked\ index, summary, serving-policy and microbenchmark files used in the analysis were verified against the manifest.

An earlier TCP-based campaign, used for comparison in Section~\ref{sec:transport}, ran in July 2026. NCCL used its default TCP socket transport on one interface. This campaign used the NGC 26.04 container (vLLM 0.19), 1{,}500\,s measurement windows and concurrency levels up to 128.


\section{Results}
\label{sec:results}

\begin{figure*}[!tp]
\centering
\includegraphics[width=\textwidth]{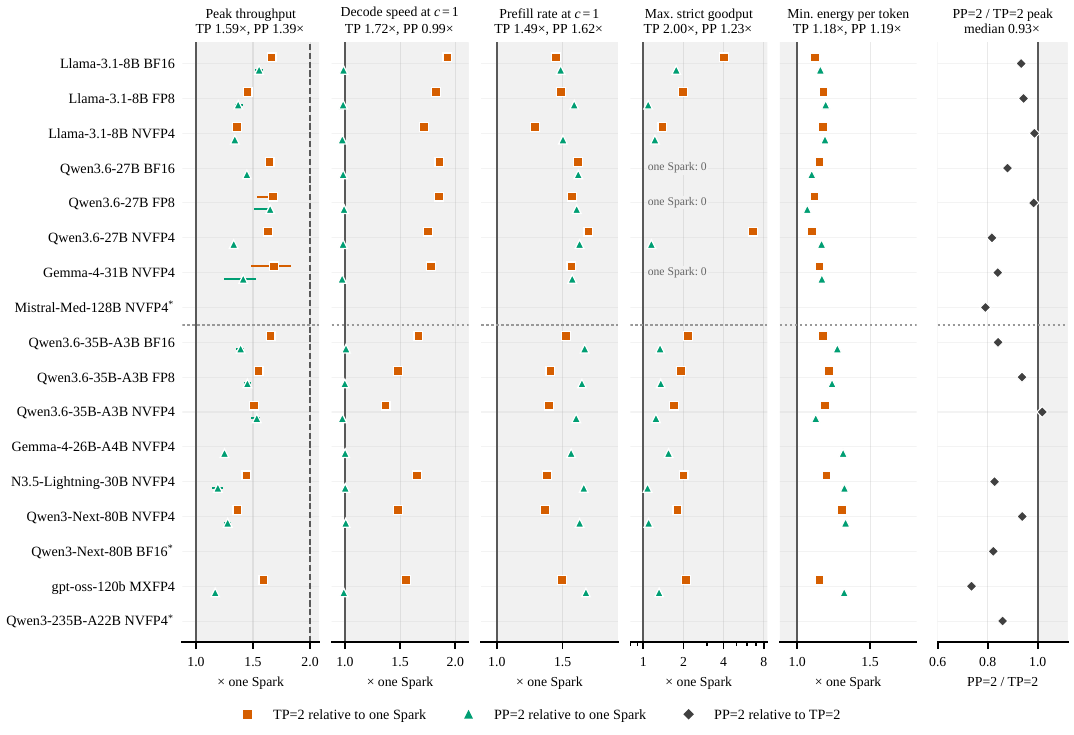}
\caption{Performance comparison of one- and two-Spark deployments across model checkpoints. The first five columns show two-Spark performance relative to one Spark: peak throughput over $c\le64$, unloaded decode speedup at $c{=}1$ (one-Spark ITL divided by two-Spark ITL), prefill rate at $c{=}1$, maximum strict-tier goodput and minimum GPU energy per token. The sixth column compares \pp\ and \tp\ peak throughput and includes the \dgxNumDualOnlyWord\ checkpoints evaluated only on two Sparks (asterisks). Dashed reference lines indicate parity, except for throughput, where the line marks ideal $2\times$ scaling. Shading indicates favorable performance regions. Goodput uses a logarithmic scale, with zero-baseline cases labeled. Whiskers in the first two columns show conservative replicate bounds, using the smallest numerator over the largest denominator and vice versa. Column headings report medians over the \dgxNumComplete\ complete checkpoints. The dotted horizontal line separates dense and MoE models.}
\label{fig:overview}
\end{figure*}

\begin{table*}[!tp]
\caption{Per-checkpoint inference performance across deployments. Peak throughput is the maximum replicate-mean output throughput over $c\le64$. Peak ratios compare two-Spark deployments against one Spark and \pp\ against \tp. ITL is the median request-mean inter-token latency at $c{=}1$, while strict-tier goodput is the maximum across tested concurrency levels. Summary medians are calculated over the \dgxNumComplete\ complete checkpoints.}
\label{tab:headline}
\centering
\footnotesize
\setlength{\tabcolsep}{3.2pt}
\begin{tabular}{@{}l r r r r r r r r r r r r@{}}
\toprule
 & \multicolumn{3}{c}{Peak tok/s} & \multicolumn{3}{c}{Peak ratio} & \multicolumn{3}{c}{ITL at $c{=}1$ (ms)} & \multicolumn{3}{c}{\shortstack{Max.\ strict\\goodput (req/s)}} \\
\cmidrule(lr){2-4}\cmidrule(lr){5-7}\cmidrule(lr){8-10}\cmidrule(lr){11-13}
Checkpoint & $1\times$ & TP & PP & TP/$1\times$ & PP/$1\times$ & PP/TP & $1\times$ & TP & PP & $1\times$ & TP & PP \\
\midrule
Llama-3.1-8B BF16 & 131 & 218 & 204 & 1.66 & 1.55 & 0.93 & 72 & 38 & 73 & 0.18 & 0.73 & 0.32 \\
Llama-3.1-8B FP8 & 321 & 466 & 439 & 1.45 & 1.37 & 0.94 & 41 & 22 & 42 & 0.79 & 1.56 & 0.85 \\
Llama-3.1-8B NVFP4 & 390 & 529 & 522 & 1.36 & 1.34 & 0.99 & 26 & 15 & 26 & 1.33 & 1.85 & 1.62 \\
Qwen3.6-27B BF16 & 64.2 & 106 & 92.8 & 1.65 & 1.45 & 0.88 & 229 & 123 & 233 & 0.00 & 0.00 & 0.00 \\
Qwen3.6-27B FP8 & 81.6 & 137 & 135 & 1.68 & 1.65 & 0.98 & 130 & 70 & 131 & 0.00 & 0.13 & 0.00 \\
Qwen3.6-27B NVFP4 & 81.1 & 132 & 108 & 1.63 & 1.33 & 0.82 & 84 & 48 & 86 & 0.03 & 0.22 & 0.04 \\
Gemma-4-31B NVFP4 & 93.7 & 158 & 133 & 1.68 & 1.41 & 0.84 & 150 & 84 & 154 & 0.00 & 0.08 & 0.00 \\
Mistral-Medium-3.5-128B NVFP4 & -- & 65.1 & 51.5 & -- & -- & 0.79 & -- & 215 & 423 & -- & 0.00 & 0.00 \\
Qwen3.6-35B-A3B BF16 & 154 & 254 & 214 & 1.65 & 1.39 & 0.84 & 34 & 20 & 33 & 0.31 & 0.68 & 0.42 \\
Qwen3.6-35B-A3B FP8 & 193 & 299 & 280 & 1.55 & 1.45 & 0.94 & 20 & 14 & 20 & 0.45 & 0.86 & 0.60 \\
Qwen3.6-35B-A3B NVFP4 & 244 & 368 & 374 & 1.51 & 1.53 & 1.02 & 14 & 9.9 & 14 & 0.84 & 1.43 & 1.04 \\
Gemma-4-26B-A4B NVFP4 & 381 & -- & 476 & -- & 1.25 & -- & 35 & -- & 35 & 0.88 & -- & 1.35 \\
Nemotron-3.5-Lightning-30B-A3B NVFP4 & 344 & 495 & 410 & 1.44 & 1.19 & 0.83 & 22 & 13 & 22 & 0.80 & 1.61 & 0.86 \\
Qwen3-Next-80B-A3B NVFP4 & 284 & 386 & 363 & 1.36 & 1.28 & 0.94 & 27 & 18 & 26 & 0.69 & 1.25 & 0.76 \\
Qwen3-Next-80B-A3B BF16 & -- & 215 & 177 & -- & -- & 0.82 & -- & 24 & 39 & -- & 0.43 & 0.27 \\
gpt-oss-120b MXFP4 & 186 & 296 & 217 & 1.59 & 1.17 & 0.74 & 28 & 18 & 28 & 0.33 & 0.69 & 0.44 \\
Qwen3-235B-A22B NVFP4 & -- & 124 & 107 & -- & -- & 0.86 & -- & 60 & 91 & -- & 0.07 & 0.04 \\
\midrule
Median (13 complete) & & & & 1.59 & 1.39 & 0.93 & & & & & & \\
\bottomrule
\end{tabular}

\end{table*}

\begin{figure*}[!tp]
\centering
\includegraphics[width=\textwidth]{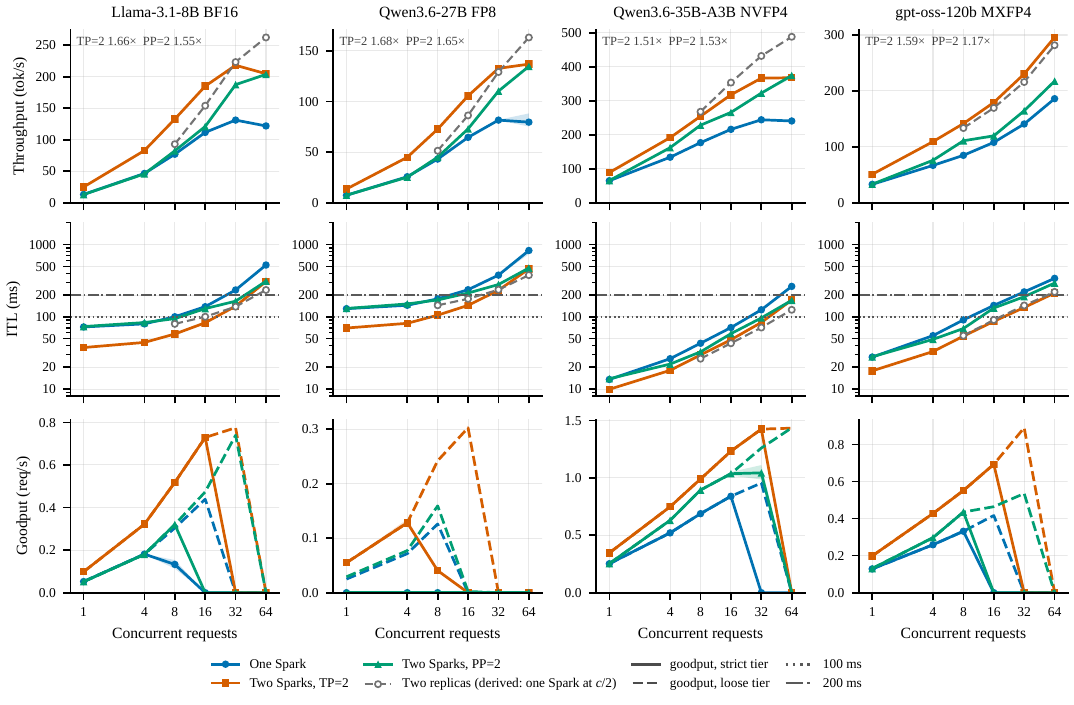}
\caption{Inference performance across concurrency levels for four representative checkpoints spanning dense and MoE architectures, four weight precisions and the extremes of the \pp/\tp\ peak throughput ratio. Rows show output throughput, median request-mean ITL on a logarithmic scale with strict (\dgxStrictSLOms\,ms) and loose (\dgxLooseSLOms\,ms) thresholds and goodput under the strict (solid) and loose (dashed) tiers. Lines show means across three independent launches, with bands indicating replicate ranges. Grey dashed lines with open markers represent two independent one-Spark replicas, derived from single-Spark measurements at half the total concurrency (Section~\ref{sec:replicas}). Annotations report peak throughput ratios relative to one Spark. Results for all checkpoints appear in Appendix~\ref{app:results}.}
\label{fig:sweeps}
\end{figure*}

Fig.~\ref{fig:overview} and Table~\ref{tab:headline} summarize the campaign and the first five subsections follow the columns of Fig.~\ref{fig:overview}. Fig.~\ref{fig:sweeps} shows the load sweeps of four representative checkpoints and Appendix~\ref{app:results} shows all \dgxNumCheckpoints.

\subsection{Throughput}
\label{sec:throughput}

Two Sparks under \tp\ increase peak throughput across all complete checkpoints by \dgxTPRatioMin--\dgxTPRatioMax\x\ (median \dgxTPRatioMed\x) relative to one Spark, compared with \dgxPPRatioMin--\dgxPPRatioMax\x\ (median \dgxPPRatioMed\x) under \pp. \tp\ achieves the higher peak on \dgxTPBestPeakCount\ of the \dgxNumComplete\ checkpoints. The exception is Qwen3.6-35B-A3B NVFP4, where \pp\ reaches \dgxQwenMoeNVPPTPRatio\x\ the \tp\ peak, although their difference at $c{=}64$ is undecided. Across \dgxCellsTPPP\ matched checkpoint and concurrency comparisons, Welch intervals place \tp\ above \pp\ in \dgxCellsThrTPgtPP\ cases, with none favoring \pp. The remaining \dgxCellsThrTPPPUnsepWord\ comparisons are undecided, all at $c{=}\dgxCellsThrTPPPUnsepLoads$. The peak throughput ratio of \pp\ to \tp\ ranges from \dgxPPTPRatioMin--\dgxPPTPRatioMax\x\ (median \dgxPPTPRatioMed\x), with the largest gap on gpt-oss-120b (\dgxGptOssPPTPRatio\x).

The difference between parallel strategies is most pronounced at low concurrency (Fig.~\ref{fig:sweeps}, top row). For Llama-3.1-8B BF16 at $c{=}1$, \tp\ achieves \dgxLlamaBFTPThrOne\,tok/s compared with \dgxLlamaBFSingleThrOne\,tok/s on one Spark, as both GPUs read their respective weight partitions concurrently. In contrast, \pp\ delivers \dgxLlamaBFPPThrOne\,tok/s, approximately matching one Spark because its stages execute sequentially for a single request. As concurrency increases, pipeline stages can overlap execution across requests, improving throughput. However, one Spark remains separated above \pp\ in \dgxCellsThrSinglegtPPWord\ of \dgxCellsSingle\ comparisons, all at $c\le\dgxCellsThrSinglegtPPMaxLoad$.

Throughput continues increasing through $c{=}64$ for \dgxCurvesPeakAtTop\ of \dgxNumCurves\ deployment curves. The exceptions are \dgxFallingSingleCountWord\ one-Spark curves (\dgxFallingSingleNames) and \dgxFallingTPCountWord\ \tp\ curve (\dgxFallingTPNames), which peak at $c{=}32$. No \pp\ curve peaks before $c{=}64$. These declines are not attributable to KV-cache exhaustion: no requests were preempted and maximum observed cache utilization was \dgxKVUsageMaxPct\,\% (\dgxKVUsageMaxModel\ \dgxKVUsageMaxConfig\ at $c{=}\dgxKVUsageMaxLoad$). Neither parallel strategy achieves twice the one-Spark peak throughput within the tested range, with the largest observed gain reaching \dgxTPRatioMax\x.

The narrowing throughput gap between \tp\ and \pp\ at higher concurrency is consistent with reduced pipeline bubbles. At low concurrency, \pp\ leaves one stage idle while the other executes, limiting GPU utilization. As more requests enter the pipeline, the stages can process different batches concurrently, reducing idle time and improving throughput. In contrast, \tp\ distributes each layer across both GPUs even at $c{=}1$, providing an advantage at low load. Although pipeline utilization improves with concurrency, \tp\ retains higher throughput in most measured configurations.

\finding{Over RDMA, \tp\ increases peak throughput across all complete checkpoints and is never separated below \pp\ at any tested concurrency. \pp\ provides smaller gains, primarily as concurrency increases enough to overlap pipeline stages.}

\subsection{Decode latency}
\label{sec:decode}

At $c{=}1$, inter-token latency (ITL) approximates the time of a single decode step without batching. \tp\ reduces ITL across all \dgxTPLowerITLCount\ complete checkpoints to \dgxTPITLRatioMin--\dgxTPITLRatioMax\x\ of the one-Spark baseline (median \dgxTPITLRatioMed\x), with every difference separated. The improvement depends on model architecture. Dense checkpoints achieve \dgxTPITLRatioDenseMin--\dgxTPITLRatioDenseMax\x, approaching the expected benefit of halving per-node weight traffic. MoE checkpoints achieve \dgxTPITLRatioMoeMin--\dgxTPITLRatioMoeMax\x, as their smaller active weight footprint makes communication and other overheads relatively more significant (Section~\ref{sec:levers}). Similarly, the benefit decreases as Llama-3.1-8B weights are quantized, with ITL ratios of \dgxLlamaBFTPITLRatio, \dgxLlamaFPTPITLRatio\ and \dgxLlamaNVTPITLRatio\x\ at BF16, FP8 and NVFP4, respectively.

In contrast, \pp\ leaves unloaded ITL nearly unchanged at \dgxPPITLRatioMin--\dgxPPITLRatioMax\x\ of one Spark, with a maximum penalty of \dgxPPITLPenaltyMaxMs\,ms. This ordering persists under load (Figs.~\ref{fig:sweeps} and \ref{fig:itl}). Across \dgxCellsTPPP\ matched checkpoint and concurrency comparisons, \tp\ has separated lower ITL in \dgxCellsITLTPlower, while \pp\ leads in none. The remaining \dgxCellsITLTPPPUnsepWord\ are undecided. At higher concurrency, \pp\ can nevertheless reduce ITL relative to one Spark by distributing requests across pipeline stages. For example, at $c{=}64$, Llama-3.1-8B BF16 achieves \dgxLlamaBFPPITLSixtyFour\,ms under \pp\ compared with \dgxLlamaBFSingleITLSixtyFour\,ms on one Spark, despite no improvement in single-request decode latency.

Unloaded ITL on one Spark ranges from \dgxITLOneMin\,ms (\dgxITLOneMinModel) to \dgxITLOneMax\,ms (\dgxITLOneMaxModel). Three dense checkpoints exceed the \dgxStrictSLOms\,ms strict-tier threshold: Qwen3.6-27B at BF16 and FP8 and Gemma-4-31B. \tp\ brings Qwen3.6-27B FP8 (\dgxQwenDenseFPTPITLOne\,ms) and Gemma-4-31B (\dgxGemmaDenseTPITLOne\,ms) below this threshold. Qwen3.6-27B BF16 remains above it under all deployments, reaching \dgxQwenDenseBFTPITLOne\,ms under \tp.

\finding{\tp\ reduces unloaded ITL to \dgxTPITLRatioDenseMin--\dgxTPITLRatioDenseMax\x\ of one Spark on dense checkpoints and \dgxTPITLRatioMoeMin--\dgxTPITLRatioMoeMax\x\ on MoE checkpoints. It consistently outperforms \pp, which provides negligible unloaded latency improvement.}

\subsection{Prefill}
\label{sec:prefill}

\begin{figure}[!tp]
\centering
\includegraphics[width=\columnwidth]{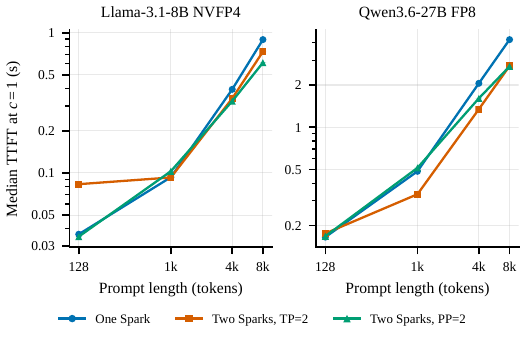}
\caption{Median time to first token (TTFT) versus prompt length at $c{=}1$ for two dense checkpoints. Points show means across three launches, with whiskers indicating replicate ranges. \pp\ reduces TTFT for prompts spanning multiple prefill chunks, while \tp\ accelerates long prompts but introduces overhead for short prompts on Llama-3.1-8B.}
\label{fig:prefill}
\end{figure}

At $c{=}1$, TTFT scales approximately linearly with prompt length across the four workload buckets, with $R^2\geq\dgxPrefillRsqMin$ for every curve. We estimate prefill rate as the reciprocal of the fitted slope. Relative to one Spark, \tp\ increases prefill rate by \dgxTPPrefillRatioMin--\dgxTPPrefillRatioMax\x\ (median \dgxTPPrefillRatioMed\x), while \pp\ achieves \dgxPPPrefillRatioMin--\dgxPPPrefillRatioMax\x\ (median \dgxPPPrefillRatioMed\x). Prefill is the only metric for which \pp\ achieves a larger median gain.

The two strategies improve prefill through different mechanisms (Fig.~\ref{fig:prefill}). Under \pp, TTFT remains near the one-Spark baseline for short prompts, at \dgxPPBucketRatioShortMin--\dgxPPBucketRatioShortMax\x\ for 128 tokens and \dgxPPBucketRatioOneKMin--\dgxPPBucketRatioOneKMax\x\ for 1{,}024 tokens. For longer prompts, it falls to \dgxPPBucketRatioFourKMin--\dgxPPBucketRatioFourKMax\x\ at 4{,}096 tokens and \dgxPPBucketRatioEightKMin--\dgxPPBucketRatioEightKMax\x\ at 8{,}192 tokens, with little variation across checkpoints. This behavior is consistent with pipeline stages overlapping work on different prefill chunks, as described in SARATHI \cite{agrawal2023sarathi}. Under an idealized two-stage pipeline, where each stage takes half the single-Spark processing time per chunk, a prompt divided into $k$ chunks completes in $(k+1)/(2k)$ of the single-Spark time. This predicts ratios of 1, 0.75 and 0.625 for one, two and four chunks, respectively. With vLLM's 2{,}048-token budget (Section~\ref{sec:setup}), these correspond closely to the observed prompt-length dependence.

Under \tp, both GPUs process each prefill chunk concurrently, reducing TTFT for 8{,}192-token prompts to \dgxTPBucketRatioEightKMin--\dgxTPBucketRatioEightKMax\x\ of the one-Spark baseline. However, collective communication can outweigh the computational savings for short prompts. On Llama-3.1-8B, \tp\ adds \dgxLlamaTPShortAddMin--\dgxLlamaTPShortAddMax\,ms for 128-token prompts. At NVFP4, TTFT increases from \dgxLlamaNVSingleBucketTTFTShort\,ms on one Spark to \dgxLlamaNVTPBucketTTFTShort\,ms under \tp. NCCL microbenchmarks provide a possible explanation: a 128-token prompt with hidden size 4096 produces 1\,MiB all-reduce payloads, requiring \dgxRdmaARPlateauMinUs--\dgxRdmaARPlateauMaxUs\,$\mu$s per collective across the measured 1--8\,MiB range (Fig.~\ref{fig:interconnect}). For Llama-3.1-8B, 64 such collectives imply approximately \dgxARPrefillShortCostMs\,ms of communication, comparable to its entire one-Spark 128-token prefill time (\dgxLlamaNVSingleBucketTTFTShort--\dgxLlamaBFSingleBucketTTFTShort\,ms across precisions).

Across the full workload mix at $c{=}1$, \tp\ reduces median TTFT to \dgxTPTTFTRatioMin--\dgxTPTTFTRatioMax\x\ of one Spark, while \pp\ remains at \dgxPPTTFTRatioMin--\dgxPPTTFTRatioMax\x. The median request contains a 1{,}024-token prompt, which does not benefit from pipeline overlap. At $c{=}64$, both strategies reduce median TTFT to approximately two-thirds of the one-Spark value (median \dgxTPTTFTAtSixtyFourRatioMed\x\ for \tp\ and \dgxPPTTFTAtSixtyFourRatioMed\x\ for \pp, Fig.~\ref{fig:ttft}).

\finding{\pp\ improves prefill primarily for prompts spanning multiple chunks, where pipeline stages can overlap execution. \tp\ accelerates long-prompt prefill but can increase TTFT for short prompts because of collective communication overhead.}

\subsection{Goodput}
\label{sec:goodput}

Goodput captures the tradeoff between throughput and latency (Fig.~\ref{fig:sweeps}, bottom row and Fig.~\ref{fig:goodput}). Every deployment that reaches the strict tier exhibits a maximum followed by a decline to zero at higher concurrency. At $c{=}64$, none of the \dgxNumCurves\ curves achieves nonzero strict-tier goodput. At the first concurrency level where goodput vanishes, the median ITL ranges from \dgxStrictCliffITLMin--\dgxStrictCliffITLMax\,ms across the \dgxStrictCliffCurves\ qualifying curves, while median TTFT exceeds \dgxTTFTSLOs\,s in only \dgxStrictCliffTTFTCountWord\ cases. Thus, the goodput collapse is primarily associated with ITL violations.

\tp\ achieves the highest maximum strict-tier goodput on all \dgxTPBestStrictCount\ complete checkpoints that reach the objective. Relative to one Spark, its gain ranges from \dgxTPStrictRatioMin--\dgxTPStrictRatioMax\x\ (median \dgxTPStrictRatioMed\x), compared with \dgxPPStrictRatioMin--\dgxPPStrictRatioMax\x\ (median \dgxPPStrictRatioMed\x) for \pp, considering checkpoints with nonzero one-Spark goodput. Two checkpoints (\dgxStrictOnlyTPNames) reach the strict tier only under \tp. This advantage follows from its lower decode latency (Section~\ref{sec:decode}), which allows the median ITL to remain below \dgxStrictSLOms\,ms for at least one additional concurrency level relative to \pp\ on \dgxStrictRangeTPgtPPCount\ of \dgxNumComplete\ checkpoints and relative to one Spark on \dgxStrictRangeTPgtSingleCount. \tp\ also achieves the highest loose-tier goodput on \dgxTPBestLooseCount\ of \dgxNumComplete\ checkpoints.

At the concurrency where \tp\ reaches its maximum goodput, its advantage over \pp\ is separated on \dgxStrictSepSameLoadCount\ of \dgxStrictNonzeroCount\ checkpoints. The remaining two (\dgxStrictUnsepSameLoadNames) exhibit high variability because \pp\ replicates straddle the \dgxStrictSLOms\,ms ITL threshold. Comparing each deployment at its own measured optimum yields separated margins on all \dgxStrictSepOwnBestCount\ checkpoints.

For models that fit on one Spark, we also compare against two derived independent replicas. Under ideal load balancing, their maximum goodput is twice the one-Spark maximum, since all single-node maxima occur at $c\le32$. Among checkpoints with nonzero one-Spark goodput, \tp\ is separated above replication on \dgxStrictRepAboveCountWord\ of \dgxStrictNonzeroSingleCountWord\ checkpoints (\dgxStrictRepAboveNames), below on \dgxStrictRepBelowCountWord\ and undecided on \dgxStrictRepUndecidedCountWord\ (\dgxStrictRepUndecidedNames). \tp\ additionally enables nonzero strict-tier goodput on two checkpoints where replication cannot meet the objective. Mistral-Medium-3.5-128B meets neither tier under either split, with unloaded ITL of \dgxMistralTPITLOne\,ms under \tp\ and \dgxMistralPPITLOne\,ms under \pp.

\finding{\tp\ achieves the highest strict-tier goodput wherever both splits are supported and the objective is attainable. It also enables \dgxStrictOnlyTPCountWord\ dense checkpoints to meet the strict tier. Compared with two derived replicas, \tp\ leads on \dgxStrictRepAheadAllCountWord\ of \dgxStrictNonzeroCountWord\ qualifying checkpoints, trails on \dgxStrictRepBelowCountWord\ and remains undecided on \dgxStrictRepUndecidedCountWord.}

\subsection{Energy and power}
\label{sec:energy}

\begin{figure*}[!tp]
\centering
\includegraphics[width=\textwidth]{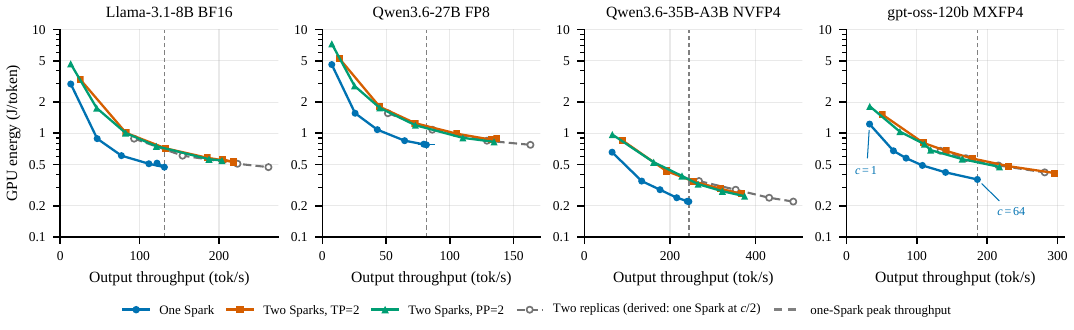}
\caption{GPU energy per generated token versus output throughput for the checkpoints in Fig.~\ref{fig:sweeps}. Each point represents one concurrency level, connected in order from $c{=}1$ to $c{=}64$, with whiskers showing replicate ranges. Grey dashed curves represent two derived one-Spark replicas, providing twice the single-Spark throughput at the same per-node concurrency and energy per token. Vertical dashed lines mark single-Spark peak throughput. One Spark provides the lowest GPU energy per token wherever its throughput is sufficient.}
\label{fig:energycurves}
\end{figure*}

GPU energy per generated token generally decreases with concurrency as throughput grows faster than power consumption (Fig.~\ref{fig:energycurves}). From $c{=}1$ to $c{=}64$, single-Spark throughput increases by \dgxSingleThrGrowthMin--\dgxSingleThrGrowthMax\x, while GPU power increases by only \dgxSinglePowerGrowthMin--\dgxSinglePowerGrowthMax\x. Energy per token decreases through $c{=}32$ on every curve, although \dgxEnergyRiseCurvesWord\ curves show a slight increase at $c{=}64$. At matched concurrency, one Spark consumes less GPU energy per token than either two-Spark deployment across all \dgxCellsEnergySingleLowerThanTP\ comparisons with \tp\ and \dgxCellsEnergySingleLowerThanPP\ comparisons with \pp, with every margin separated.

At $c{=}64$, \tp\ consumes a median \dgxEnergyTPAtSixtyFourMed\x\ the single-Spark energy per token (range \dgxEnergyTPAtSixtyFourMin--\dgxEnergyTPAtSixtyFourMax\x), compared with \dgxEnergyPPAtSixtyFourMed\x\ under \pp\ (range \dgxEnergyPPAtSixtyFourMin--\dgxEnergyPPAtSixtyFourMax\x). The difference is more pronounced at $c{=}1$: \tp\ consumes a median \dgxEnergyTPAtOneMed\x\ the single-Spark energy per token, compared with \dgxEnergyPPAtOneMed\x\ for \pp. Tensor parallelism offsets some of the additional GPU power through faster decoding, whereas pipeline parallelism provides little unloaded latency improvement. Across \dgxCellsTPPP\ matched comparisons, \tp\ has separated lower energy per token in \dgxCellsEnergyTPLowerThanPP, while \pp\ leads in \dgxCellsEnergyPPLowerThanTPWord, all at $c{=}64$. The remaining \dgxCellsEnergyTPPPUnsepWord\ comparisons are undecided. Nevertheless, \pp\ achieves a lower minimum energy per token on \dgxEnergyMinPPBelowTPCountWord\ checkpoints (Table~\ref{tab:energy}).

Minimum energy per token varies substantially across checkpoints, from \dgxEnergyFloorMin\,J for \dgxEnergyFloorMinModel\ on one Spark to \dgxEnergyFloorMax\,J for \dgxEnergyFloorMaxModel\ under \pp, a \dgxEnergyFloorSpread\x\ difference (Fig.~\ref{fig:energygrid} and Table~\ref{tab:energy}). These measurements include GPU energy reported by \texttt{nvidia-smi}, summed across both nodes for distributed deployments. CPU, networking and power-supply energy are excluded.

\finding{One Spark minimizes GPU energy per generated token at matched concurrency, while two-Spark deployments provide higher throughput at an energy cost. At $c{=}64$, \tp\ and \pp\ consume a median \dgxEnergyTPAtSixtyFourMed\x\ and \dgxEnergyPPAtSixtyFourMed\x\ the single-Spark energy per token, respectively.}

\subsection{Splitting against replicating}
\label{sec:replicas}

Models that fit on one Spark can alternatively be deployed as two independent replicas, with requests distributed between them. We derive this baseline from single-Spark measurements, assuming perfect load balancing without routing overhead, independent nodes and identical performance on both Sparks. At total concurrency $c$, the replica pair therefore achieves twice the one-Spark throughput at $c/2$, while retaining the corresponding latency and GPU energy per token. This comparison is available for total concurrency from \dgxRepLoadMin\ to 64. Figs.~\ref{fig:sweeps} and \ref{fig:energycurves} show the derived baseline alongside the measured parallel deployments.

The preferred strategy changes with concurrency. At $c{=}8$, \tp\ achieves higher throughput than replication on \dgxRepThrTPAboveEight\ of \dgxNumComplete\ complete checkpoints, delivering \dgxRepThrTPRatioEightMin--\dgxRepThrTPRatioEightMax\x\ of the replica pair's throughput across all checkpoints. It also achieves lower ITL on \dgxRepITLTPLowerEight\ checkpoints and lower GPU energy per token on \dgxRepEnergyTPLowerEight. At $c{=}64$, replication delivers higher throughput on \dgxRepThrTPBelowSixtyFour\ checkpoints, with \tp\ achieving \dgxRepThrTPRatioSixtyFourMin--\dgxRepThrTPRatioSixtyFourMax\x\ of the replica baseline. Replication also achieves lower ITL on \dgxRepITLTPHigherSixtyFour\ checkpoints and lower energy per token on \dgxRepEnergyTPHigherSixtyFour. \tp\ remains ahead in throughput only on \dgxRepThrTPAboveSixtyFourNames, while \dgxRepThrTPUndecidedSixtyFourNames\ is undecided.

\pp\ is never separated above replication in throughput, latency, or energy efficiency at any evaluated load, reaching at most \dgxRepThrPPRatioMax\x\ of the replica pair's throughput. These results reflect the tradeoff between accelerating individual requests and serving independent batches. At low concurrency, \tp\ reduces decode latency by distributing each layer across both GPUs. At high concurrency, replication processes separate batches without the per-layer synchronization required by \tp.

\finding{For models that fit on one Spark, \tp\ generally favors low-concurrency serving, while independent replication offers higher throughput and lower latency at high concurrency. \pp\ provides no separated advantage over replication in the evaluated metrics.}

\subsection{Checkpoints run only on two Sparks}
\label{sec:dualonly}

Three checkpoints were evaluated only on two Sparks. Qwen3-Next-80B-A3B BF16 and Qwen3-235B-A22B NVFP4 exceed the memory capacity of one Spark, while Mistral-Medium-3.5-128B NVFP4 requires \dgxMistralTotalGiB\,GiB for weights, leaving limited KV-cache capacity within the configured \dgxGPUUtil\ memory fraction.

Across these checkpoints, \pp\ achieves \dgxDualOnlyPPTPRatioMin--\dgxDualOnlyPPTPRatioMax\x\ of the \tp\ peak throughput, with \dgxDualOnlyPPTPITLRatioMin--\dgxDualOnlyPPTPITLRatioMax\x\ its unloaded ITL (Table~\ref{tab:headline}). The largest throughput gap occurs on Mistral-Medium-3.5-128B, where \tp\ delivers \dgxMistralTPPeak\,tok/s compared with \dgxMistralPPPeak\,tok/s under \pp, with unloaded ITL of \dgxMistralTPITLOne\ and \dgxMistralPPITLOne\,ms, respectively. Under \tp, both GPUs process their weight partitions concurrently, whereas \pp\ executes its stages sequentially at $c{=}1$. Of the two checkpoints reaching the strict-tier objective, \tp\ achieves higher maximum goodput, with a separated margin on \dgxStrictDualOnlySepNames\ and an undecided margin on \dgxStrictDualOnlyUndecidedNames. Thus, \tp\ is the preferred split for all three checkpoints based on throughput and decode latency.

Beyond supporting larger models, two Sparks increase KV-cache capacity by \dgxTPKVRatioMin--\dgxTPKVRatioMax\x\ under \tp\ and \dgxPPKVRatioMin--\dgxPPKVRatioMax\x\ under \pp\ across the \dgxNumComplete\ complete checkpoints (Table~\ref{tab:models}). Most gains exceed twofold because model weights are partitioned across nodes, leaving more aggregate memory for the KV cache. This additional capacity was not a limiting factor in our workload, where maximum cache utilization reached only \dgxKVUsageMaxPct\,\% (Section~\ref{sec:throughput}). Workloads with longer contexts or higher concurrency could benefit more substantially.

\subsection{Model-side levers}
\label{sec:levers}

\begin{figure*}[!tp]
\centering
\includegraphics[width=\textwidth]{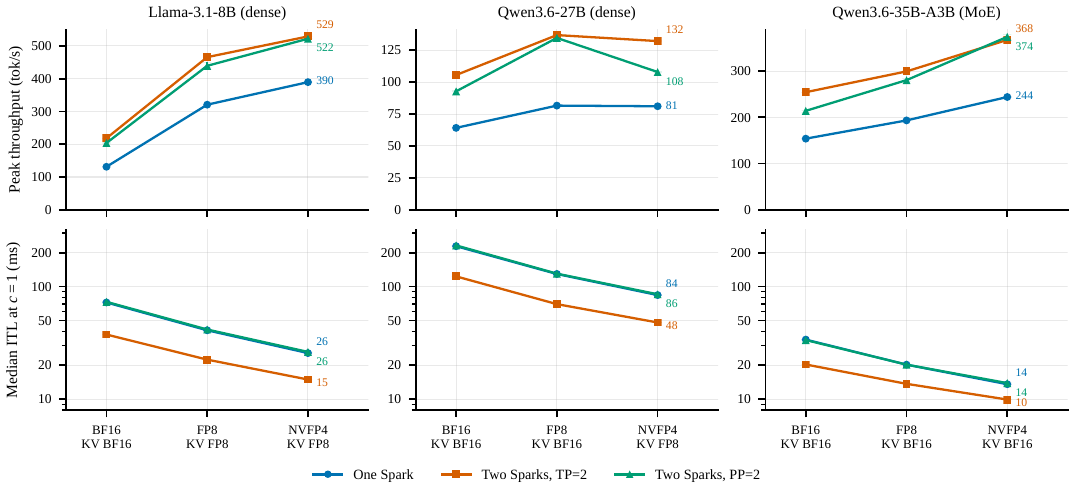}
\caption{Effect of weight precision across three model families at BF16, FP8 and NVFP4. Top: peak output throughput. Bottom: unloaded median request-mean ITL at $c{=}1$. Tick labels indicate the served KV-cache precision, which changes across some ModelOpt checkpoints. The Qwen3.6 dense and MoE models also provide an architectural comparison at matched weight precision.}
\label{fig:precision}
\end{figure*}

\begin{figure}[!tp]
\centering
\includegraphics[width=\columnwidth]{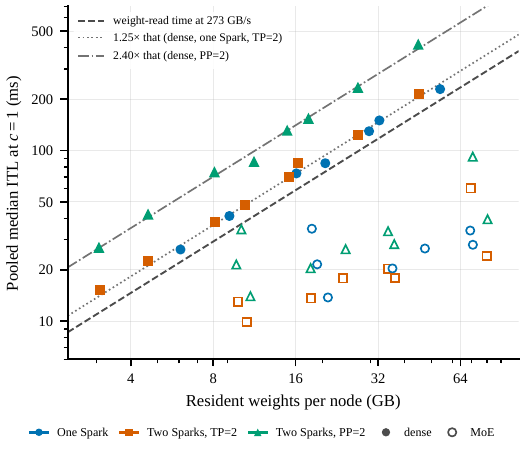}
\caption{Unloaded decode latency (pooled median token gap) versus resident weight footprint per node. The dashed line represents the time to read the weights once at the nominal \dgxSpecBW\,GB/s memory bandwidth. Dotted and dash-dotted lines scale this reference by the median factors observed for dense models under one Spark and \tp\ (\dgxBoundDenseDirectMed\x) and \pp\ (\dgxBoundDensePPMed\x), respectively. MoE checkpoints (open markers) fall below the full-weight reference because only active experts are accessed during decoding.}
\label{fig:bound}
\end{figure}

\textbf{Precision.}
Fig.~\ref{fig:precision} compares three model families at BF16, FP8 and NVFP4. Across all three, reducing precision from BF16 to NVFP4 on one Spark lowers unloaded ITL to \dgxLadderITLRatioMin--\dgxLadderITLRatioMax\x\ of the BF16 value, a larger improvement than adding a second Spark under \tp\ at BF16 (\dgxLadderBFTPITLRatioMin--\dgxLadderBFTPITLRatioMax\x). The impact on peak throughput varies by model. Quantization improves throughput by \dgxLadderLlamaPeakGain\x\ for Llama-3.1-8B, \dgxLadderQwenMoePeakGain\x\ for Qwen3.6-35B-A3B and \dgxLadderQwenDensePeakGain\x\ for Qwen3.6-27B, compared with respective BF16 \tp\ gains of \dgxLlamaBFTPRatio, \dgxQwenMoeBFTPRatio\ and \dgxQwenDenseBFTPRatio\x. The benefits are complementary: at NVFP4, \tp\ provides additional throughput gains of \dgxLlamaNVTPRatio, \dgxQwenMoeNVTPRatio\ and \dgxQwenDenseNVTPRatio\x, respectively.

Qwen3.6-27B is an exception to the expected throughput improvement from lower precision. Its NVFP4 checkpoint achieves \dgxQwenDenseNVSinglePeak\,tok/s, comparable to FP8 (\dgxQwenDenseFPSinglePeak\,tok/s), despite reducing unloaded ITL from \dgxQwenDenseFPSingleITLOne\ to \dgxQwenDenseNVSingleITLOne\,ms. Its prefill rate also falls to \dgxQwenDenseNVSinglePrefill\ input tokens/s, compared with \dgxQwenDenseFPSinglePrefill\ for FP8 and \dgxQwenDenseBFSinglePrefill\ for BF16. The checkpoint uses weight-only NVFP4 (\texttt{W4A16\_NVFP4}), executed through vLLM's Marlin kernel with 16-bit activations (Appendix~\ref{app:config}). This execution path is consistent with improved memory-bound decoding but slower compute-intensive prefill. However, the MoE checkpoint using the same kernel path benefits from NVFP4, so kernel selection alone does not establish the cause.

Some ModelOpt checkpoints also change KV-cache precision from BF16 to FP8 (Fig.~\ref{fig:precision}). Consequently, these comparisons reflect the released checkpoints as served rather than isolated changes in weight precision.

\textbf{Sparsity.}
At matched weight precision, the MoE Qwen3.6-35B-A3B achieves \dgxQwenPairPeakRatioMin--\dgxQwenPairPeakRatioMax\x\ the peak throughput of dense Qwen3.6-27B on one Spark, with \dgxQwenPairITLRatioMin--\dgxQwenPairITLRatioMax\x\ shorter unloaded ITL. Similarly, Gemma-4-26B-A4B achieves \dgxGemmaMoeSinglePeak\,tok/s compared with \dgxGemmaDenseSinglePeak\,tok/s for Gemma-4-31B. These comparisons illustrate the serving advantages associated with sparse expert activation, although differences in model size and training prevent attributing the gains solely to sparsity.

Fig.~\ref{fig:bound} relates unloaded decode latency to memory traffic. Dense checkpoints on one Spark require \dgxBoundDenseSingleMin--\dgxBoundDenseSingleMax\x\ the theoretical time to read their weights once at \dgxSpecBW\,GB/s, corresponding to an effective weight-read bandwidth of \dgxEffBWDenseSingleMin--\dgxEffBWDenseSingleMax\,GB/s. Additional latency includes KV-cache accesses, computation and other overheads. Under \tp, each GPU reads approximately half the weights concurrently, yielding \dgxBoundDenseTPMin--\dgxBoundDenseTPMax\x\ the corresponding weight-read bound. Under \pp, stages execute sequentially at $c{=}1$, producing \dgxBoundDensePPMin--\dgxBoundDensePPMax\x\ the per-node weight-read reference. MoE checkpoints require only \dgxBoundMoeSingleMin--\dgxBoundMoeSingleMax\x\ of the full-weight reference time because only active experts are accessed. Their reduced weight traffic also explains why \tp\ provides smaller unloaded latency gains for MoE than dense checkpoints (Section~\ref{sec:decode}).

\finding{Reducing precision from BF16 to NVFP4 improves unloaded latency more than adding a second Spark under \tp\ across all three precision ladders and the benefits combine. Peak throughput gains range from \dgxLadderPeakGainMin\x\ to \dgxLadderPeakGainMax\x, with the smallest gain associated with a weight-only NVFP4 checkpoint served through the Marlin kernel.}

\subsection{Run-to-run variation}
\label{sec:variance}

\begin{figure}[!tp]
\centering
\includegraphics[width=\columnwidth]{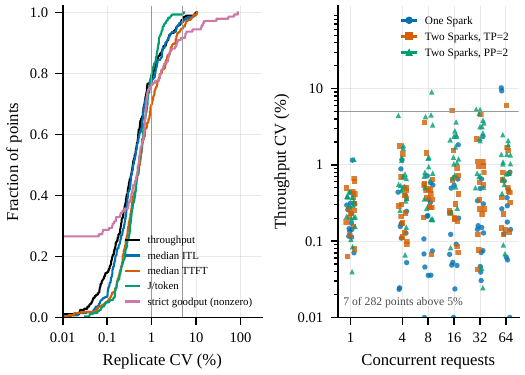}
\caption{Variability across \dgxNumPoints\ measurement points. Left: distribution of the coefficient of variation (CV) across three independent launches for each metric, with strict goodput restricted to points where it is nonzero. Right: throughput CV by concurrency and deployment, with a 5\,\% reference line.}
\label{fig:variance}
\end{figure}

\begin{table}[!tp]
\caption{Replicate coefficient of variation (CV), defined as standard deviation divided by mean, across \dgxNumPoints\ measurement points. Values report the median, 90th percentile, maximum and number of points exceeding 5\,\% CV.}
\label{tab:variance}
\centering
\footnotesize
\setlength{\tabcolsep}{4pt}
\begin{tabular}{@{}l r r r r@{}}
\toprule
Metric & Median & 90th pct. & Max & $n > 5\%$ \\
\midrule
Output throughput & 0.38 & 2.13 & 10.2 & 7 \\
Median ITL & 0.39 & 2.04 & 10.4 & 9 \\
Median TTFT & 0.54 & 3.33 & 10.6 & 15 \\
Strict goodput & 0.50 & 3.67 & 86.6 & 12 \\
Loose goodput & 0.55 & 3.86 & 173.2 & 17 \\
Energy per token & 0.51 & 1.44 & 5.3 & 1 \\
GPU power & 0.51 & 1.61 & 8.2 & 1 \\
\bottomrule
\end{tabular}

\end{table}

Fig.~\ref{fig:variance} and Table~\ref{tab:variance} quantify run-to-run variability across three independent launches. Median CVs are \dgxCVThrMedian\,\% for throughput, \dgxCVITLMedian\,\% for ITL, \dgxCVTTFTMedian\,\% for TTFT and \dgxCVEnergyMedian\,\% for energy per token. The 90th percentile throughput CV is \dgxCVThrPNinety\,\%, with only \dgxCVThrCountOverFiveWord\ points exceeding 5\,\%. In each outlier, two launches agree within \dgxCVPairSpreadMax\,\%, while the third differs by \dgxCVOddLaunchDevMin--\dgxCVOddLaunchDevMax\,\%. Two Qwen3-235B-A22B outliers originate from the same slow launch. Two others, Gemma-4-31B NVFP4 and Qwen3.6-27B FP8 at $c{=}64$ on one Spark, exhibit one faster launch despite identical logged kernel selections. The slower launches consumed less power without higher temperatures, suggesting that thermal throttling was not responsible.

Strict-tier goodput exhibits greater variability (maximum CV \dgxCVStrictMax\,\%) because small latency differences near the ITL threshold can cause goodput to drop to zero. We therefore report confidence intervals for goodput comparisons (Section~\ref{sec:goodput}). Despite this variability, the main throughput ordering remains consistent: \tp\ is separated above one Spark in all \dgxCellsThrTPgtSingle\ of \dgxCellsSingle\ matched comparisons, while \pp\ is separated above it in \dgxCellsThrPPgtSingle\ and below it in \dgxCellsThrSinglegtPPWord, with the latter occurring only at $c\le\dgxCellsThrSinglegtPPMaxLoad$.


\section{The Interconnect: From TCP to RDMA}
\label{sec:transport}

Our initial DGX Spark experiments used NCCL over TCP sockets (\texttt{NET/Socket}). At the time, the out-of-box software configuration did not automatically provision the ConnectX-7 RoCEv2 network for our workload and NCCL consequently fell back to TCP. NVIDIA subsequently introduced updated NCCL support and the NVIDIA Sync Cluster Assistant to simplify multi-node network configuration. In the present campaign, we configured all four RoCEv2 rails and verified that every dual-node launch used NCCL's RDMA transport (\texttt{NET/IB}, Section~\ref{sec:setup}).

The earlier TCP campaign favored \pp\ over \tp, consistent with conventional multi-node guidance. In contrast, \tp\ outperforms \pp\ in the present RDMA campaign. We compare the two campaigns and use NCCL microbenchmarks to investigate how reduced communication overhead contributes to this reversal.

\begin{figure}[!tp]
\centering
\includegraphics[width=\columnwidth]{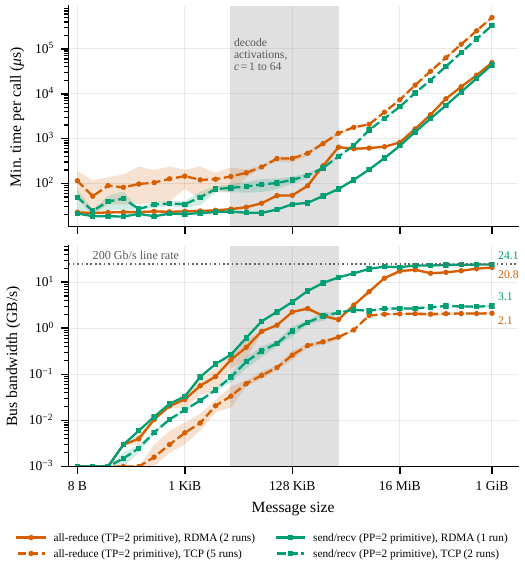}
\caption{NCCL communication performance over RDMA (solid) and TCP sockets (dashed). Top: minimum operation latency across message sizes. Bottom: effective bus bandwidth. Orange curves show all-reduce, the primary \tp\ communication primitive, and green curves show send/recv, used by \pp. Bands indicate variation across runs. The shaded region marks decode-size payloads from 8\,KiB to 1\,MiB. RDMA substantially reduces all-reduce overhead in this range. TCP measurements used one cable and a different worker node.}
\label{fig:interconnect}
\end{figure}

\subsection{Communication overhead}

Fig.~\ref{fig:interconnect} compares the communication primitives used by both parallel strategies. Within the 8\,KiB to 1\,MiB decode payload range, RDMA provides \dgxBandARRdmaOverTcpMin--\dgxBandARRdmaOverTcpMax\x\ the all-reduce bandwidth of TCP. For an 8\,KiB message, the mean minimum all-reduce latency falls from \dgxTcpARLatEightKiBMin\,$\mu$s over TCP to \dgxRdmaARLatEightKiBMin\,$\mu$s over RDMA.

This reduction is particularly important for \tp, which requires approximately two all-reduces per transformer layer during decoding. Llama-3.1-8B performs \dgxLlamaAllReducesPerStep\ such operations per step, corresponding to an estimated collective cost of \dgxARStepCostTcpMs\,ms over TCP versus \dgxARStepCostRdmaMs\,ms over RDMA. The estimated saving of \dgxARStepSavingMs\,ms per token is comparable to the observed reduction of \dgxXLlamaBFTPITLDropMs, \dgxXLlamaFPTPITLDropMs\ and \dgxXLlamaNVTPITLDropMs\,ms across its BF16, FP8 and NVFP4 checkpoints.

Although send/recv remains more communication-efficient than all-reduce, \tp\ benefits from both GPUs reading their weight partitions concurrently, whereas \pp\ processes its stages sequentially during single-request decoding (Section~\ref{sec:levers}). With lower collective overhead, this parallel memory access advantage becomes more pronounced.

\begin{figure*}[!tp]
\centering
\includegraphics[width=\textwidth]{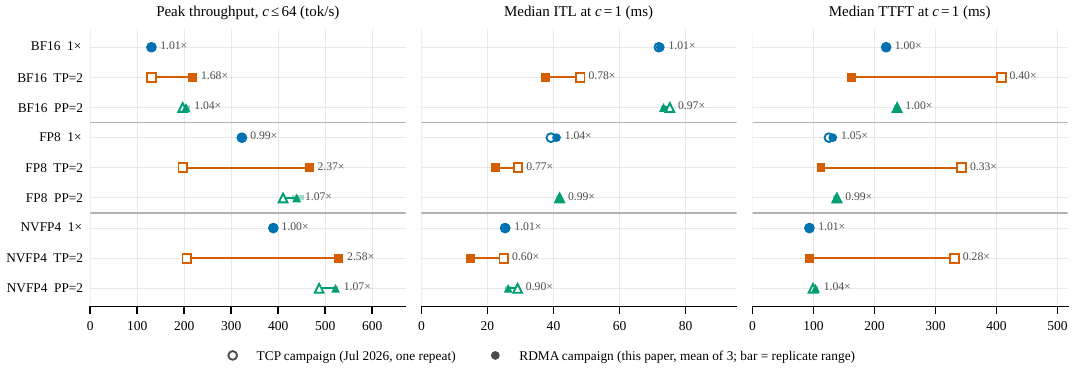}
\caption{Performance comparison of the July 2026 TCP campaign (open markers) and the present RDMA campaign (filled markers) across three Llama-3.1-8B precision variants. Columns show peak throughput, unloaded ITL and TTFT at $c{=}1$. Annotations give RDMA-to-TCP ratios. One-Spark results provide a control unaffected by inter-node communication. The campaigns also differ in software versions, worker node, cabling and measurement methodology.}
\label{fig:crosscampaign}
\end{figure*}

\subsection{Cross-campaign comparison}

Fig.~\ref{fig:crosscampaign} compares the three Llama-3.1-8B checkpoints measured in both campaigns. Relative to TCP, \tp\ peak throughput increases by \dgxXLlamaBFTPPeakRatio\x\ at BF16, \dgxXLlamaFPTPPeakRatio\x\ at FP8 and \dgxXLlamaNVTPPeakRatio\x\ at NVFP4. In contrast, \pp\ improves by only \dgxXLlamaBFPPPeakRatio--\dgxXLlamaFPPPPeakRatio\x, while one-Spark throughput remains within approximately 1\,\% of its earlier value.

The largest relative improvement occurs in \tp\ time to first token. At $c{=}1$, TTFT decreases by \dgxXLlamaTPTTFTDropMsMin--\dgxXLlamaTPTTFTDropMsMax\,ms, reaching \dgxXLlamaTPTTFTRatioMin--\dgxXLlamaTPTTFTRatioMax\x\ of its TCP value. For the median 1{,}024-token prompt, the NCCL microbenchmarks predict approximately \dgxARPrefillOneKSavingMs\,ms of collective communication savings across the 64 all-reduces, accounting for most of the observed reduction. These results are consistent with communication overhead being a major factor in the reversal between \tp\ and \pp. 

\finding{Between the TCP and RDMA campaigns, \tp\ peak throughput increases by \dgxXLlamaBFTPPeakRatio--\dgxXLlamaNVTPPeakRatio\x, substantially more than \pp. NCCL microbenchmarks show that reduced collective latency can explain much of the improvement in \tp\ inference performance, supporting the interconnect as a major contributor to the reversal in parallelism performance.}


\section{Discussion}
\label{sec:discussion}

\subsection{Choosing a deployment}

Our measurements suggest the following deployment guidelines for RDMA-connected DGX Spark systems running the evaluated vLLM configuration.

\textbf{Models requiring two Sparks: Prefer \tp.}
For the \dgxNumDualOnlyWord\ checkpoints that could not be served on one Spark under our configuration, \tp\ delivers higher peak throughput and lower unloaded latency. Across these checkpoints, \pp\ achieves only \dgxDualOnlyPPTPRatioMin--\dgxDualOnlyPPTPRatioMax\x\ of \tp's peak throughput, with \dgxDualOnlyPPTPITLRatioMin--\dgxDualOnlyPPTPITLRatioMax\x\ its unloaded ITL (Section~\ref{sec:dualonly}).

\textbf{Latency-sensitive or low-concurrency workloads: Prefer \tp.}
For models that fit on one Spark, \tp\ reduces unloaded ITL to \dgxTPITLRatioMin--\dgxTPITLRatioMax\x\ of the single-Spark baseline. Unlike replication, which increases aggregate serving capacity without accelerating individual requests, \tp\ improves single-request decoding. At low concurrency, it also outperforms two derived replicas in throughput on most checkpoints (Section~\ref{sec:replicas}). For \dgxStrictOnlyTPNames, this latency improvement enables strict-tier goodput that neither one Spark nor \pp\ achieves.

\textbf{High-concurrency workloads: Prefer replication.}
When the model fits on one Spark, independent replicas generally provide better throughput at high concurrency. At $c{=}64$, the derived replica pair outperforms \tp\ on \dgxRepThrTPBelowSixtyFourWord\ of the \dgxNumCompleteWord\ complete checkpoints, generally with lower latency and GPU energy per token. Replication also provides higher throughput than \pp\ across the evaluated checkpoints. These comparisons assume ideal load balancing and identical node performance.

\textbf{Energy-constrained workloads: Consider one Spark.}
A single Spark consistently consumes less GPU energy per generated token at matched concurrency than either distributed deployment (Section~\ref{sec:energy}). When its throughput and latency satisfy the workload requirements, adding a second Spark offers no GPU-energy-efficiency advantage. Total system energy, however, was not measured.

\textbf{When to consider \pp.}
Pipeline parallelism offers limited advantages in our workload. For models fitting on one Spark, it never achieves a separated advantage over the derived replica pair in throughput, latency, or energy efficiency. For larger models, \tp\ generally performs better. However, \pp\ provides faster prefill for sufficiently long prompts: at 8{,}192 tokens and $c{=}1$, it achieves lower TTFT than \tp\ on \dgxPPBeatsTPEightKCountWord\ of the \dgxNumCompleteWord\ complete checkpoints. These measurements contain relatively few long-prompt requests. Thus, workloads dominated by long prompts at low concurrency, or models lacking \tp\ kernel support, may favor \pp.

\textbf{Quantize first, when model quality permits.}
Across the three precision ladders, moving from BF16 to NVFP4 reduces unloaded latency more than adding a second Spark under \tp\ and the improvements are complementary (Section~\ref{sec:levers}). However, quantization does not guarantee higher throughput. Qwen3.6-27B NVFP4 illustrates how weight-only execution through the Marlin kernel can improve decode latency while limiting prefill and throughput gains. Precision variants should therefore be benchmarked under the intended workload before deployment.

\subsection{Limitations}

\textbf{Scope.}
The study evaluates two GB10 systems, pipeline depth two and a single vLLM version. The synthetic workload uses fixed-length greedy decoding, closed-loop concurrency, disabled prefix caching and approximately \dgxMixInOutRatio\ prompt tokens per generated token. Different workload mixes may change the relative benefits of prefill and decode optimizations. Model quality under quantization was not evaluated.

\textbf{Derived replication baseline.}
Two independent replicas were modeled using single-Spark measurements, assuming identical node performance, perfect load balancing and no routing overhead. Actual replica performance may differ.

\textbf{Measurement and generalizability.}
Concurrency was limited to $c{=}64$, where most throughput curves were still increasing. Reported peaks therefore represent maxima within the tested range rather than saturation throughput. Statistical comparisons use three independent launches per deployment without correction for multiple testing. GPU energy measurements exclude CPU, networking and other system components. NCCL transfers were staged through host memory because GPUDirect RDMA is unsupported on GB10.

\textbf{Checkpoint coverage.}
Two measured checkpoints were excluded from the analysis (Section~\ref{sec:checkpoints}) and \tp\ could not be evaluated for Gemma-4-26B-A4B due to unsupported kernels in the tested vLLM version.


\section{Related Work}
\label{sec:related}

\textbf{Serving systems and parallelism.}
vLLM introduced PagedAttention for efficient KV-cache management \cite{kwon2023vllm}, building on Orca's iteration-level scheduling \cite{yu2022orca}. SARATHI introduced chunked prefill \cite{agrawal2023sarathi}, while DistServe formalized goodput as a serving objective \cite{zhong2024distserve}. Megatron-LM \cite{shoeybi2019megatron} and GPipe \cite{huang2019gpipe} established tensor and pipeline parallelism, with subsequent work recommending tensor parallelism within servers and pipeline parallelism across servers \cite{narayanan2021}. Pope et al.\ examine latency-throughput tradeoffs in inference partitioning \cite{pope2023efficiently}, while Xu et al.\ characterize distributed inference communication patterns \cite{xu2025characterizingcommunicationpatternsdistributed}. Our work extends these investigations to desktop-scale GPU systems with unified memory and RDMA-connected nodes, where inter-node communication and memory bandwidth affect the relative benefits of parallelism strategies.

\textbf{Inference benchmarking.}
MLPerf Inference establishes standardized performance scenarios and accuracy requirements \cite{reddi2020mlperf}. LLM-Inference-Bench evaluates inference performance across accelerators and serving frameworks \cite{chittyvenkata2024llmbench}, while Etalon highlights the limitations of aggregate latency metrics for interactive serving \cite{agrawal2024etalon}. Samsi et al.\ investigate LLM inference energy consumption on datacenter GPUs \cite{samsi2023watts}. We complement these studies by evaluating throughput, latency, SLO-based goodput and GPU energy efficiency across replicated single- and dual-node experiments.

\textbf{DGX Spark evaluations.}
Existing evaluations include single-node performance studies \cite{lmsys2025spark,vllm2026spark} and a two-Spark cluster review \cite{storagereview2026sparkcluster}. The cluster review compares \tp\ and \pp\ on Llama-3.1-8B and gpt-oss-120b using a single 200\,GbE connection. For gpt-oss-120b, it reports \pp\ achieving 2.20\x\ the \tp\ throughput at batch size 128 with equal input and output lengths, whereas \tp\ leads at lower batch sizes on Llama-3.1-8B. These findings differ from our RDMA results, where \tp\ leads on gpt-oss-120b throughout the evaluated concurrency range. However, differences in workloads, concurrency, software configurations and unspecified network transport prevent direct attribution of this discrepancy. Our study extends prior evaluations through a broader checkpoint matrix, three independent server launches per configuration, verified NCCL transport and comparisons of model splitting with derived independent replication, including goodput and GPU energy measurements.


\section{Conclusion}
\label{sec:conclusion}

We systematically evaluated LLM inference on one and two RDMA-connected DGX Spark GB10 systems across \dgxNumCheckpoints\ model checkpoints. Contrary to conventional multi-node guidance, tensor parallelism (\tp) consistently outperforms pipeline parallelism (\pp) in throughput, decode latency and goodput. Across the \dgxNumComplete\ complete checkpoints, \tp\ increases median peak throughput to \dgxTPRatioMed\x\ that of one Spark and reduces median unloaded inter-token latency to \dgxTPITLRatioMed\x\ of the single-node baseline. In contrast, \pp\ achieves a median throughput gain of \dgxPPRatioMed\x\ without improving unloaded decode latency, although it accelerates prefill for sufficiently long prompts.

The preferred deployment depends on workload characteristics. For models requiring two Sparks or workloads sensitive to decode latency, \tp\ provides the strongest overall performance. For models that fit on one Spark, \tp\ often outperforms two derived independent replicas at low concurrency, while replication delivers higher throughput on \dgxRepThrTPBelowSixtyFourWord\ of the \dgxNumCompleteWord\ complete checkpoints at $c{=}64$. A single Spark remains the most GPU-energy-efficient configuration at matched concurrency. Weight quantization provides an additional, complementary means of improving inference performance.

Comparisons with an earlier TCP-based campaign and NCCL microbenchmarks suggest that reduced collective communication overhead contributes substantially to \tp's advantage over \pp. Although other configuration changes prevent isolating RDMA's contribution, the results demonstrate that parallelism strategies should be evaluated against the memory architecture, interconnect and workload rather than selected solely from conventional datacenter guidance.

\section*{Acknowledgments}

This work was conducted at the First State AI Institute (FSAII) at the University of Delaware. This material is based upon work supported by the National Science Foundation under Grant No. 2417814 through the Democratizing Access to Research Software Engineers (DARSE) project. The authors thank ASUS for generously providing two DGX Spark systems used in this research.

Anthropic's Claude was used to assist with code development and manuscript editing, including language polishing and improvements to clarity. All scientific claims, experimental results and manuscript content were thoroughly reviewed and verified by the authors, who retain full responsibility for the final work.

\bibliographystyle{IEEEtran}
\bibliography{refs}

@misc{nvidia2025sparkhw,
  author       = {{NVIDIA Corporation}},
  title        = {{DGX} {Spark} User Guide: Hardware Overview},
  year         = {2025},
  howpublished = {\url{https://docs.nvidia.com/dgx/dgx-spark/hardware.html}},
  note         = {Accessed: 2026-11-13}
}

@misc{nvidia2025sparkstack,
  author       = {{NVIDIA Corporation}},
  title        = {{DGX} {Spark} User Guide: {ConnectX-7} Networking},
  year         = {2025},
  howpublished = {\url{https://docs.nvidia.com/dgx/dgx-spark/spark-clustering.html}},
  note         = {Accessed: 2026-11-13}
}

@inproceedings{kwon2023vllm,
  author    = {Kwon, Woosuk and Li, Zhuohan and Zhuang, Siyuan and Sheng, Ying and Zheng, Lianmin and Yu, Cody Hao and Gonzalez, Joseph E. and Zhang, Hao and Stoica, Ion},
  title     = {Efficient Memory Management for Large Language Model Serving with {PagedAttention}},
  booktitle = {Proceedings of the 29th Symposium on Operating Systems Principles (SOSP)},
  pages     = {611--626},
  doi       = {10.1145/3600006.3613165},
  year      = {2023}
}

@inproceedings{yu2022orca,
  author    = {Yu, Gyeong-In and Jeong, Joo Seong and Kim, Geon-Woo and Kim, Soojeong and Chun, Byung-Gon},
  title     = {Orca: A Distributed Serving System for Transformer-Based Generative Models},
  booktitle = {16th USENIX Symposium on Operating Systems Design and Implementation (OSDI)},
  pages     = {521--538},
  year      = {2022}
}

@inproceedings{zhong2024distserve,
  author    = {Zhong, Yinmin and Liu, Shengyu and Chen, Junda and Hu, Jianbo and Zhu, Yibo and Liu, Xuanzhe and Jin, Xin and Zhang, Hao},
  title     = {{DistServe}: Disaggregating Prefill and Decoding for Goodput-optimized Large Language Model Serving},
  booktitle = {18th USENIX Symposium on Operating Systems Design and Implementation (OSDI)},
  pages     = {193--210},
  year      = {2024}
}

@inproceedings{moritz2018ray,
  author    = {Moritz, Philipp and Nishihara, Robert and Wang, Stephanie and Tumanov, Alexey and Liaw, Richard and Liang, Eric and Elibol, Melih and Yang, Zongheng and Paul, William and Jordan, Michael I. and Stoica, Ion},
  title     = {Ray: A Distributed Framework for Emerging {AI} Applications},
  booktitle = {13th USENIX Symposium on Operating Systems Design and Implementation (OSDI)},
  pages     = {561--577},
  year      = {2018}
}

@article{shoeybi2019megatron,
  author  = {Shoeybi, Mohammad and Patwary, Mostofa and Puri, Raul and LeGresley, Patrick and Casper, Jared and Catanzaro, Bryan},
  title   = {{Megatron-LM}: Training Multi-Billion Parameter Language Models Using Model Parallelism},
  journal = {arXiv preprint arXiv:1909.08053},
  year    = {2019}
}

@inproceedings{huang2019gpipe,
  author    = {Huang, Yanping and Cheng, Youlong and Bapna, Ankur and Firat, Orhan and Chen, Mia Xu and Chen, Dehao and Lee, HyoukJoong and Ngiam, Jiquan and Le, Quoc V. and Wu, Yonghui and Chen, Zhifeng},
  title     = {{GPipe}: Efficient Training of Giant Neural Networks using Pipeline Parallelism},
  booktitle = {Advances in Neural Information Processing Systems 32 (NeurIPS)},
  year      = {2019}
}

@inproceedings{narayanan2021,
  author    = {Narayanan, Deepak and Shoeybi, Mohammad and Casper, Jared and LeGresley, Patrick and Patwary, Mostofa and Korthikanti, Vijay and Vainbrand, Dmitri and Kashinkunti, Prethvi and Bernauer, Julie and Catanzaro, Bryan and Phanishayee, Amar and Zaharia, Matei},
  title     = {Efficient Large-Scale Language Model Training on {GPU} Clusters Using {Megatron-LM}},
  booktitle = {Proceedings of the International Conference for High Performance Computing, Networking, Storage and Analysis (SC)},
  series    = {SC '21},
  publisher = {Association for Computing Machinery},
  address   = {New York, NY, USA},
  doi       = {10.1145/3458817.3476209},
  url       = {https://doi.org/10.1145/3458817.3476209},
  articleno = {58},
  numpages  = {15},
  location  = {St. Louis, Missouri},
  isbn      = {9781450384421},
  year      = {2021}
}

@inproceedings{xu2025characterizingcommunicationpatternsdistributed,
  title     = {Characterizing Communication Patterns in Distributed Large Language Model Inference},
  author    = {Lang Xu and Kaushik Kandadi Suresh and Quentin Anthony and Nawras Alnaasan and Dhabaleswar K. Panda},
  booktitle = {2025 IEEE Symposium on High-Performance Interconnects (HOTI)},
  pages     = {1--11},
  doi       = {10.1109/HOTI66940.2025.00025},
  year      = {2025}
}

@inproceedings{pope2023efficiently,
  author    = {Pope, Reiner and Douglas, Sholto and Chowdhery, Aakanksha and Devlin, Jacob and Bradbury, James and Levskaya, Anselm and Heek, Jonathan and Xiao, Kefan and Agrawal, Shivani and Dean, Jeff},
  title     = {Efficiently Scaling Transformer Inference},
  booktitle = {Proceedings of Machine Learning and Systems 5 (MLSys)},
  pages     = {606--624},
  year      = {2023}
}

@article{williams2009roofline,
  author    = {Williams, Samuel and Waterman, Andrew and Patterson, David},
  title     = {Roofline: An Insightful Visual Performance Model for Multicore Architectures},
  journal   = {Communications of the ACM},
  volume    = {52},
  number    = {4},
  pages     = {65--76},
  month     = apr,
  year      = {2009},
  issn      = {0001-0782},
  doi       = {10.1145/1498765.1498785},
  url       = {https://doi.org/10.1145/1498765.1498785}
}

@misc{yuan2024llminferenceunveiledsurvey,
  title         = {{LLM} Inference Unveiled: Survey and Roofline Model Insights},
  author        = {Zhihang Yuan and Yuzhang Shang and Yang Zhou and Zhen Dong and Zhe Zhou and Chenhao Xue and Bingzhe Wu and Zhikai Li and Qingyi Gu and Yong Jae Lee and Yan Yan and Beidi Chen and Guangyu Sun and Kurt Keutzer},
  year          = {2024},
  eprint        = {2402.16363},
  archivePrefix = {arXiv},
  primaryClass  = {cs.CL},
  url           = {https://arxiv.org/abs/2402.16363}
}

@inproceedings{reddi2020mlperf,
  author    = {Reddi, Vijay Janapa and Cheng, Christine and Kanter, David and Mattson, Peter and others},
  title     = {{MLPerf} Inference Benchmark},
  booktitle = {ACM/IEEE 47th Annual International Symposium on Computer Architecture (ISCA)},
  pages     = {446--459},
  doi       = {10.1109/ISCA45697.2020.00045},
  year      = {2020}
}

@inproceedings{chittyvenkata2024llmbench,
  author    = {Chitty-Venkata, Krishna Teja and Raskar, Siddhisanket and Kale, Bharat and Ferdaus, Farah and Tanikanti, Aditya and Raffenetti, Ken and Taylor, Valerie and Emani, Murali and Vishwanath, Venkatram},
  title     = {{LLM-Inference-Bench}: Inference Benchmarking of Large Language Models on {AI} Accelerators},
  booktitle = {SC24-W: Workshops of the International Conference for High Performance Computing, Networking, Storage and Analysis (PMBS)},
  publisher = {IEEE},
  pages     = {1362--1379},
  doi       = {10.1109/SCW63240.2024.00178},
  year      = {2024}
}

@misc{agrawal2024etalon,
  title         = {Etalon: Holistic Performance Evaluation Framework for {LLM} Inference Systems},
  author        = {Amey Agrawal and Anmol Agarwal and Nitin Kedia and Jayashree Mohan and Souvik Kundu and Nipun Kwatra and Ramachandran Ramjee and Alexey Tumanov},
  year          = {2024},
  eprint        = {2407.07000},
  archivePrefix = {arXiv},
  primaryClass  = {cs.LG},
  url           = {https://arxiv.org/abs/2407.07000}
}

@misc{micikevicius2022fp8formatsdeeplearning,
  title         = {{FP8} Formats for Deep Learning},
  author        = {Paulius Micikevicius and Dusan Stosic and Neil Burgess and Marius Cornea and Pradeep Dubey and Richard Grisenthwaite and Sangwon Ha and Alexander Heinecke and Patrick Judd and John Kamalu and Naveen Mellempudi and Stuart Oberman and Mohammad Shoeybi and Michael Siu and Hao Wu},
  year          = {2022},
  eprint        = {2209.05433},
  archivePrefix = {arXiv},
  primaryClass  = {cs.LG},
  url           = {https://arxiv.org/abs/2209.05433}
}

@misc{alvarez2025nvfp4,
  author       = {Eduardo Alvarez and Omri Almog and Eric Chung and Simon Layton and Dusan Stosic and Ronny Krashinsky and Kyle Aubrey},
  title        = {Introducing {NVFP4} for Efficient and Accurate Low-Precision Inference},
  howpublished = {NVIDIA Technical Blog, \url{https://developer.nvidia.com/blog/introducing-nvfp4-for-efficient-and-accurate-low-precision-inference/}},
  month        = jun,
  year         = {2025},
  note         = {Accessed: 2026-11-13}
}

@misc{nvidia2025modelopt,
  author       = {{NVIDIA Corporation}},
  title        = {{NVIDIA Model Optimizer} ({ModelOpt})},
  howpublished = {\url{https://github.com/NVIDIA/Model-Optimizer}},
  year         = {2025},
  note         = {Formerly TensorRT Model Optimizer. Accessed: 2026-11-13}
}

@inproceedings{shazeer2017moe,
  author    = {Shazeer, Noam and Mirhoseini, Azalia and Maziarz, Krzysztof and Davis, Andy and Le, Quoc and Hinton, Geoffrey and Dean, Jeff},
  title     = {Outrageously Large Neural Networks: The Sparsely-Gated Mixture-of-Experts Layer},
  booktitle = {International Conference on Learning Representations (ICLR)},
  year      = {2017}
}

@article{fedus2022switch,
  author  = {Fedus, William and Zoph, Barret and Shazeer, Noam},
  title   = {Switch Transformers: Scaling to Trillion Parameter Models with Simple and Efficient Sparsity},
  journal = {Journal of Machine Learning Research},
  volume  = {23},
  number  = {120},
  pages   = {1--39},
  year    = {2022}
}

@misc{grattafiori2024llama3,
  title         = {The {Llama} 3 Herd of Models},
  author        = {Aaron Grattafiori and others},
  year          = {2024},
  eprint        = {2407.21783},
  archivePrefix = {arXiv},
  primaryClass  = {cs.AI},
  url           = {https://arxiv.org/abs/2407.21783}
}

@misc{yang2025qwen3,
  title         = {{Qwen3} Technical Report},
  author        = {An Yang and others},
  year          = {2025},
  eprint        = {2505.09388},
  archivePrefix = {arXiv},
  primaryClass  = {cs.CL},
  url           = {https://arxiv.org/abs/2505.09388}
}

@misc{nvidia2025nemotron3,
  title         = {{NVIDIA} {Nemotron} 3: Efficient and Open Intelligence},
  author        = {{NVIDIA}},
  year          = {2025},
  eprint        = {2512.20856},
  archivePrefix = {arXiv},
  primaryClass  = {cs.CL},
  url           = {https://arxiv.org/abs/2512.20856}
}

@misc{nvidia2026nemotron3super,
  title         = {{Nemotron 3 Super}: Open, Efficient Mixture-of-Experts Hybrid {Mamba-Transformer} Model for Agentic Reasoning},
  author        = {{NVIDIA}},
  year          = {2026},
  eprint        = {2604.12374},
  archivePrefix = {arXiv},
  primaryClass  = {cs.LG},
  url           = {https://arxiv.org/abs/2604.12374}
}

@misc{lmsys2025spark,
  author       = {Jerry Zhou and Richard Chen},
  title        = {{NVIDIA} {DGX} {Spark} In-Depth Review: A New Standard for Local {AI} Inference},
  howpublished = {LMSYS Org Blog, \url{https://lmsys.org/blog/2025-10-13-nvidia-dgx-spark/}},
  month        = oct,
  year         = {2025},
  note         = {Accessed: 2026-11-13}
}

@misc{vllm2026spark,
  author       = {{Inferact}},
  title        = {{vLLM} on the {DGX} {Spark}: Architecture, Configuration, and Local Evaluation},
  howpublished = {vLLM Blog, \url{https://vllm-project.github.io/2026/06/01/vllm-dgx-spark.html}},
  month        = jun,
  year         = {2026},
  note         = {Accessed: 2026-11-13}
}

@misc{storagereview2026sparkcluster,
  author       = {Dylan Dougherty and Divyansh Jain},
  title        = {{NVIDIA} {DGX} {Spark} Cluster Review: Distributed Inference on {Dell}, {GIGABYTE}, and {HP}},
  howpublished = {StorageReview, \url{https://www.storagereview.com/review/nvidia-dgx-spark-cluster-review-distributed-inference-on-dell-gigabyte-and-hp}},
  month        = may,
  year         = {2026},
  note         = {Accessed: 2026-11-13}
}

@inproceedings{samsi2023watts,
author = {Siddharth Samsi and Dan Zhao and Joseph McDonald and Baolin Li and Adam Michaleas and Michael Jones and William Bergeron and Jeremy Kepner and Devesh Tiwari and Vijay Gadepally},
title = {From Words to Watts: Benchmarking the Energy Costs of Large Language Model Inference},
booktitle = {2023 IEEE High Performance Extreme Computing Conference (HPEC)},
pages = {1--9},
doi = {10.1109/HPEC58863.2023.10363447},
year = {2023}
}

@misc{ibta2014rocev2,
  author       = {{InfiniBand Trade Association}},
  title        = {Supplement to {InfiniBand Architecture Specification Volume 1 Release 1.2.1}, {Annex A17}: {RoCEv2}},
  month        = sep,
  year         = {2014},
  howpublished = {\url{https://www.infinibandta.org/ibta-specification/}}
}

@misc{nvidia2026nccl,
  author       = {{NVIDIA Corporation}},
  title        = {{NVIDIA Deep Learning NCCL} Documentation},
  year         = {2026},
  howpublished = {\url{https://docs.nvidia.com/deeplearning/nccl/}},
  note         = {Accessed: 2026-11-13}
}

@misc{nvidia2026vllmrelnotes,
  author       = {{NVIDIA Corporation}},
  title        = {{vLLM} Release Notes, Release 26.09},
  year         = {2026},
  howpublished = {\url{https://docs.nvidia.com/deeplearning/frameworks/vllm-release-notes/rel-26-09.html}},
  note         = {Accessed: 2026-11-13}
}

@misc{nvidia2026sparkplaybook,
  author       = {{NVIDIA Corporation}},
  title        = {{DGX} {Spark} Playbooks: Connect Two {Sparks}},
  year         = {2026},
  howpublished = {\url{https://github.com/NVIDIA/dgx-spark-playbooks/tree/main/nvidia/connect-two-sparks}},
  note         = {Accessed: 2026-11-13}
}

@misc{openai2025gptoss,
  title         = {gpt-oss-120b \& gpt-oss-20b Model Card},
  author        = {{OpenAI}},
  year          = {2025},
  eprint        = {2508.10925},
  archivePrefix = {arXiv},
  primaryClass  = {cs.CL},
  url           = {https://arxiv.org/abs/2508.10925}
}

@misc{qwen2026qwen36,
  author       = {{Qwen Team}},
  title        = {{Qwen3.6-27B} and {Qwen3.6-35B-A3B} model cards},
  year         = {2026},
  howpublished = {Hugging Face, \url{https://huggingface.co/Qwen/Qwen3.6-27B} and \url{https://huggingface.co/Qwen/Qwen3.6-35B-A3B}},
  note         = {Accessed: 2026-11-13}
}

@misc{qwen2025qwen3next,
  author       = {{Qwen Team}},
  title        = {{Qwen3-Next-80B-A3B-Instruct} model card},
  year         = {2025},
  howpublished = {Hugging Face, \url{https://huggingface.co/Qwen/Qwen3-Next-80B-A3B-Instruct}},
  note         = {Accessed: 2026-11-13}
}

@misc{gemma2026report,
  title         = {{Gemma 4} Technical Report},
  author        = {{Gemma Team}},
  year          = {2026},
  eprint        = {2607.02770},
  archivePrefix = {arXiv},
  primaryClass  = {cs.CL},
  url           = {https://arxiv.org/abs/2607.02770}
}

@misc{google2026gemma4,
  author       = {{Google DeepMind}},
  title        = {{Gemma 4} model cards ({Gemma-4-31B-IT} and {Gemma-4-26B-A4B-IT})},
  year         = {2026},
  howpublished = {Hugging Face, \url{https://huggingface.co/google/gemma-4-31B-it} and \url{https://huggingface.co/google/gemma-4-26B-A4B-it}},
  note         = {Served from the NVIDIA ModelOpt NVFP4 checkpoints \url{https://huggingface.co/nvidia/Gemma-4-31B-IT-NVFP4} and \url{https://huggingface.co/nvidia/Gemma-4-26B-A4B-NVFP4}. Accessed: 2026-11-13}
}

@misc{mistral2026medium35,
  author       = {{Mistral AI}},
  title        = {{Mistral Medium 3.5 128B} model card},
  year         = {2026},
  howpublished = {Hugging Face, \url{https://huggingface.co/mistralai/Mistral-Medium-3.5-128B}; served from the NVIDIA ModelOpt NVFP4 checkpoint \url{https://huggingface.co/nvidia/Mistral-Medium-3.5-128B-NVFP4}},
  note         = {Accessed: 2026-11-13}
}

@misc{nvidia2026lightning,
  author       = {{NVIDIA}},
  title        = {{NVIDIA-Nemotron-3.5-Lightning-30B-A3B-NVFP4} model card},
  year         = {2026},
  howpublished = {Hugging Face, \url{https://huggingface.co/nvidia/NVIDIA-Nemotron-3.5-Lightning-30B-A3B-NVFP4}},
  note         = {Accessed: 2026-11-13}
}

@misc{nvidia2026qwen235nvfp4,
  author       = {{NVIDIA}},
  title        = {{Qwen3-235B-A22B-Instruct-2507-NVFP4} model card},
  year         = {2026},
  howpublished = {Hugging Face, \url{https://huggingface.co/nvidia/Qwen3-235B-A22B-Instruct-2507-NVFP4}},
  note         = {Accessed: 2026-11-13}
}

@misc{vllm2026pr57197,
  author       = {{vLLM contributors}},
  title        = {{[Bugfix][PP][Spec Decode] MiniMax-M3 EAGLE3 aux-state relay at PP $>$ 1 and per-stage FlashInfer autotune}},
  year         = {2026},
  howpublished = {vLLM pull request \#57197, \url{https://github.com/vllm-project/vllm/pull/57197}},
  note         = {Merged 2026-09-30, after the vLLM 0.29.0 release; first released in vLLM 0.32.0. Accessed: 2026-11-13}
}

@misc{vllm2026issue52291,
  author       = {{vLLM contributors}},
  title        = {{[Bug]: Synchronized FlashInfer autotune deadlocks multi-node TP startup on hosts without GPUDirect RDMA (GB10/DGX Spark)}},
  year         = {2026},
  howpublished = {vLLM issue \#52291, \url{https://github.com/vllm-project/vllm/issues/52291}},
  note         = {Opened 2026-08-14. Accessed: 2026-11-13}
}

@misc{nvidia2026sparkporting,
  author       = {{NVIDIA Corporation}},
  title        = {{NVIDIA} {DGX} {Spark} Porting Guide: {CUDA}},
  year         = {2026},
  howpublished = {\url{https://docs.nvidia.com/dgx/dgx-spark-porting-guide/latest/porting/cuda.html}},
  note         = {Accessed: 2026-11-13}
}

@misc{agrawal2023sarathi,
  title         = {{SARATHI}: Efficient {LLM} Inference by Piggybacking Decodes with Chunked Prefills},
  author        = {Amey Agrawal and Ashish Panwar and Jayashree Mohan and Nipun Kwatra and Bhargav S. Gulavani and Ramachandran Ramjee},
  year          = {2023},
  eprint        = {2308.16369},
  archivePrefix = {arXiv},
  primaryClass  = {cs.LG},
  url           = {https://arxiv.org/abs/2308.16369}
}

@article{welch1947,
  author  = {Welch, B. L.},
  title   = {The generalization of `{Student's}' problem when several different population variances are involved},
  journal = {Biometrika},
  volume  = {34},
  number  = {1--2},
  pages   = {28--35},
  year    = {1947},
  doi     = {10.1093/biomet/34.1-2.28}
}

\appendices

\section{Additional Results}
\label{app:results}

\begin{figure*}[p]
\centering
\includegraphics[width=\textwidth]{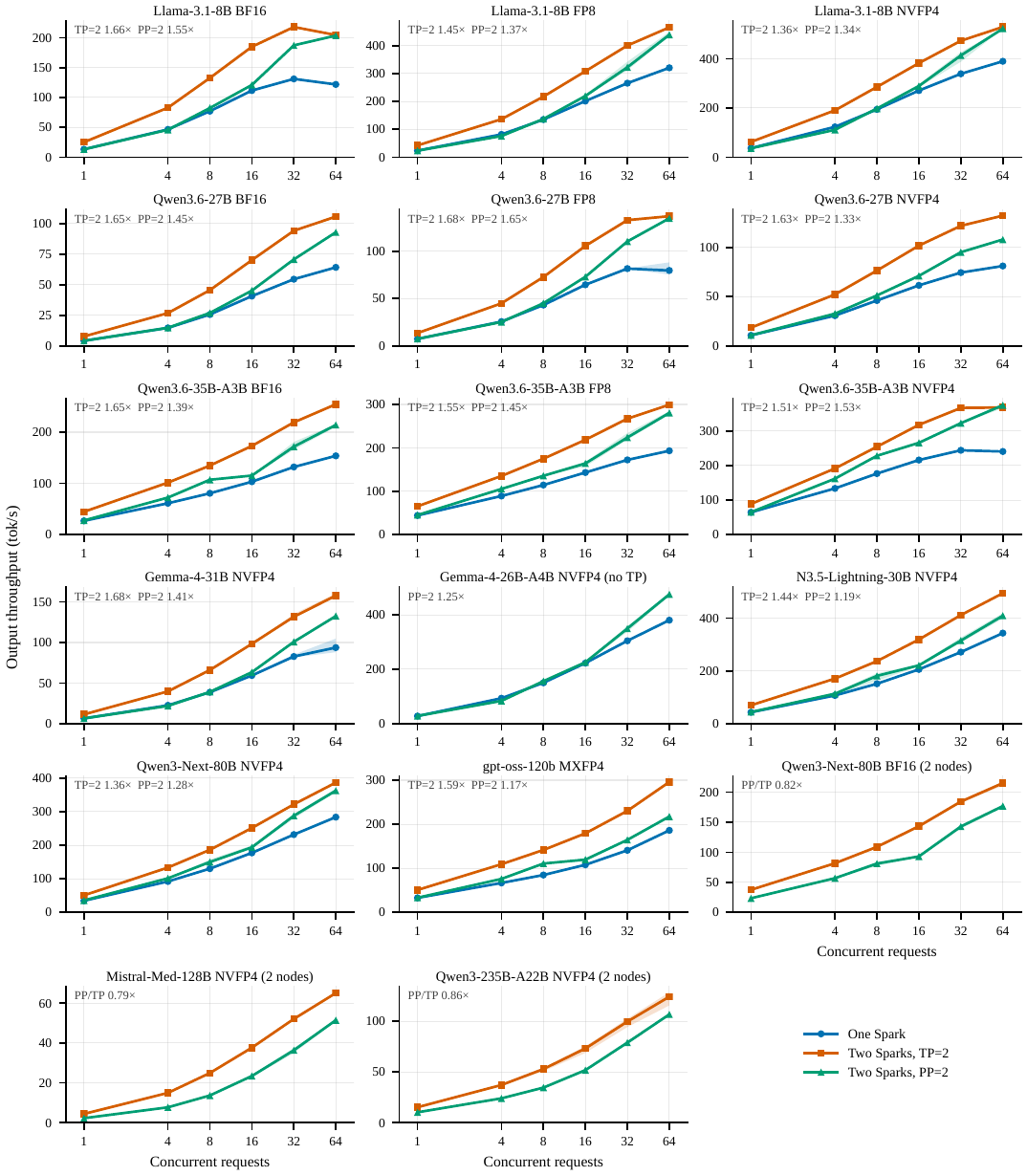}
\caption{Output throughput versus concurrency across all checkpoints. Lines show means over three independent launches, with bands indicating replicate ranges. Annotations report peak throughput relative to one Spark, or the \pp/\tp\ peak ratio for checkpoints evaluated only on two Sparks.}
\label{fig:throughput}
\end{figure*}

\begin{figure*}[p]
\centering
\includegraphics[width=\textwidth]{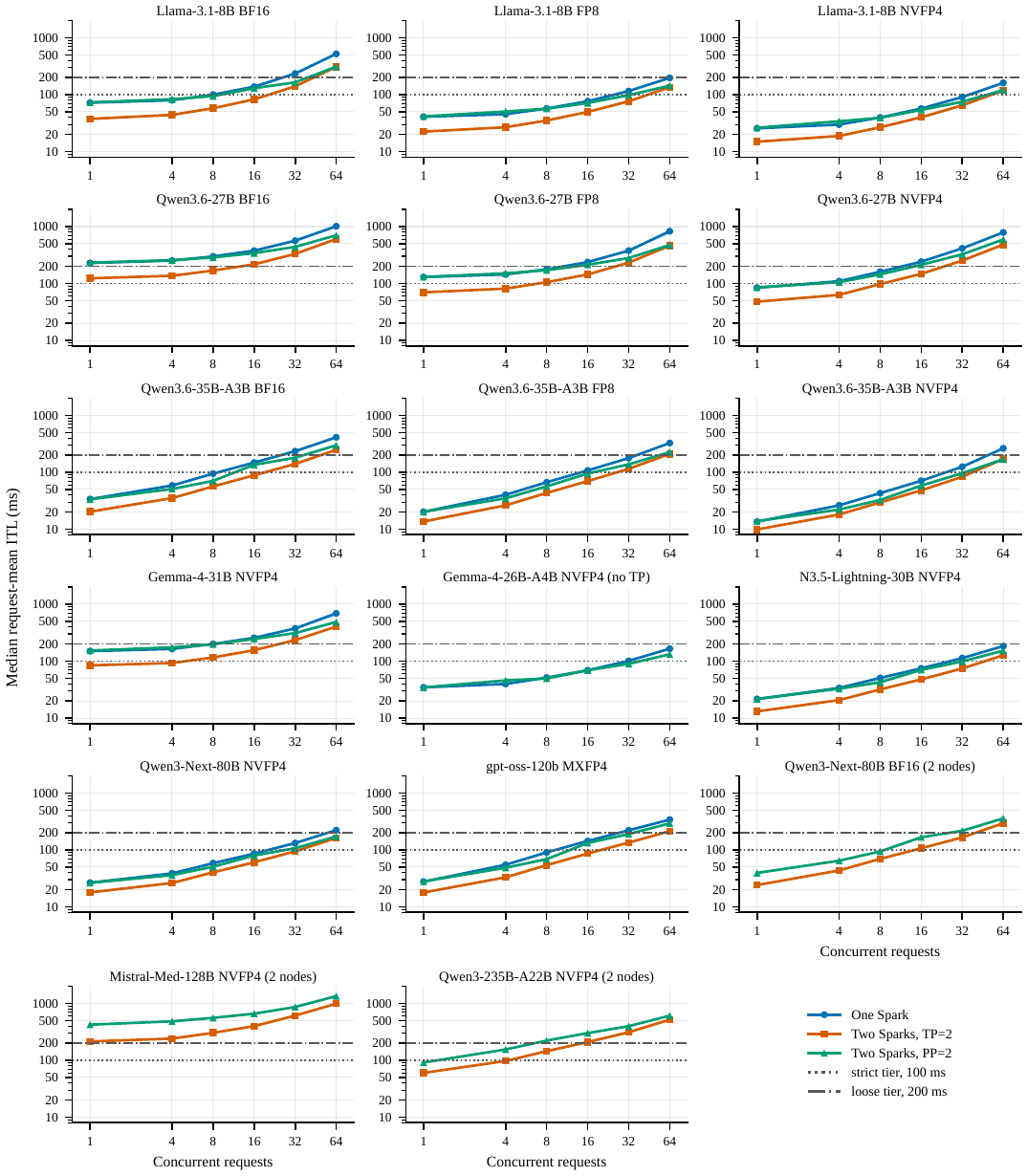}
\caption{Median request-mean inter-token latency (ITL) versus concurrency on a shared logarithmic scale. Horizontal reference lines indicate the strict (\dgxStrictSLOms\,ms) and loose (\dgxLooseSLOms\,ms) latency thresholds. Where measured, \tp\ consistently reduces ITL relative to one Spark, while \pp\ provides little improvement at low concurrency but benefits from increasing load.}
\label{fig:itl}
\end{figure*}

\begin{figure*}[p]
\centering
\includegraphics[width=\textwidth]{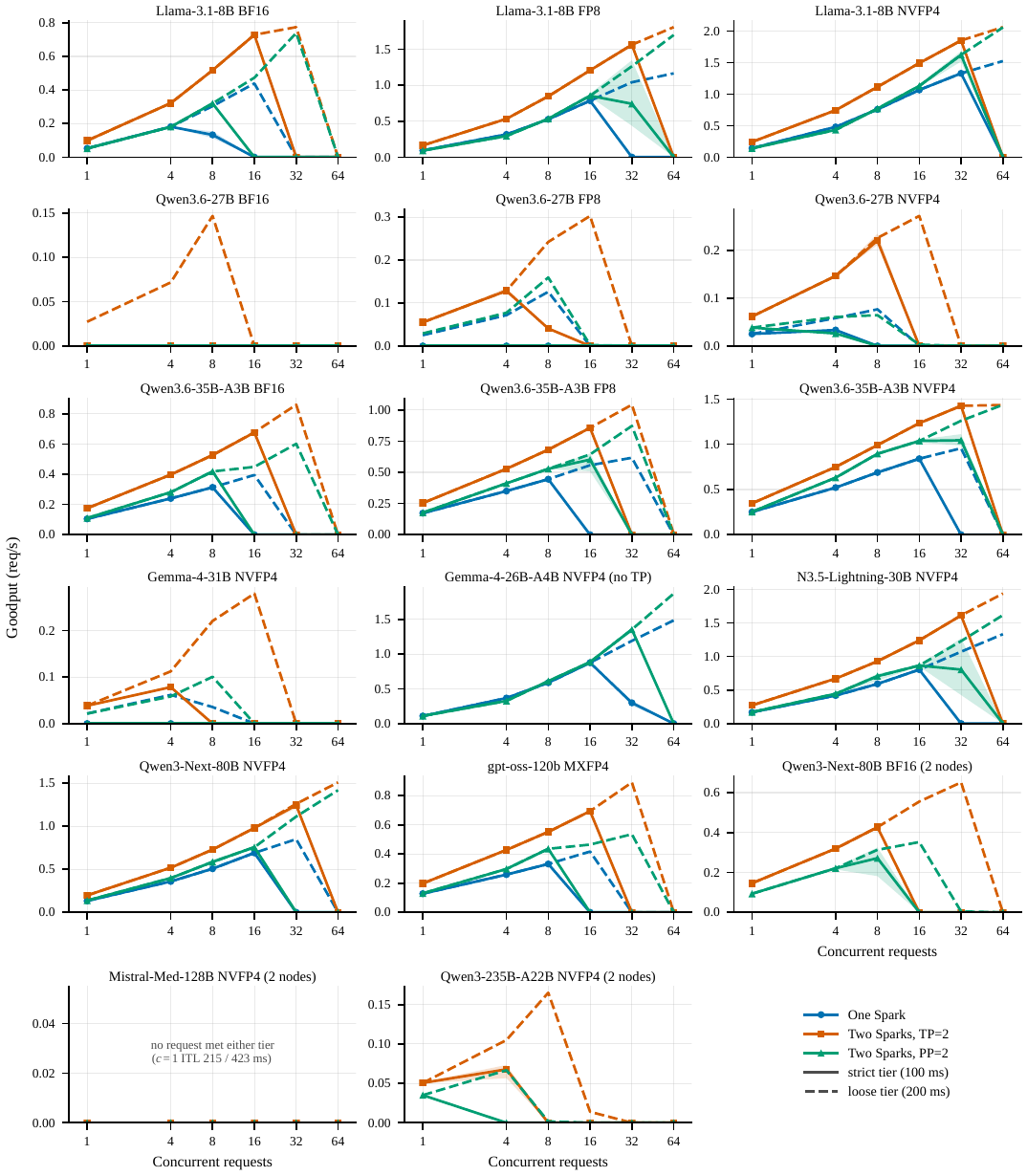}
\caption{Goodput versus concurrency under strict (\dgxStrictSLOms\,ms, solid) and loose (\dgxLooseSLOms\,ms, dashed) ITL objectives, with TTFT limited to \dgxTTFTSLOs\,s. Solid curves include replicate ranges. Zero goodput indicates that no requests satisfied both latency objectives, not that inference stopped. Panels use independent vertical scales.}
\label{fig:goodput}
\end{figure*}

\begin{figure*}[p]
\centering
\includegraphics[width=\textwidth]{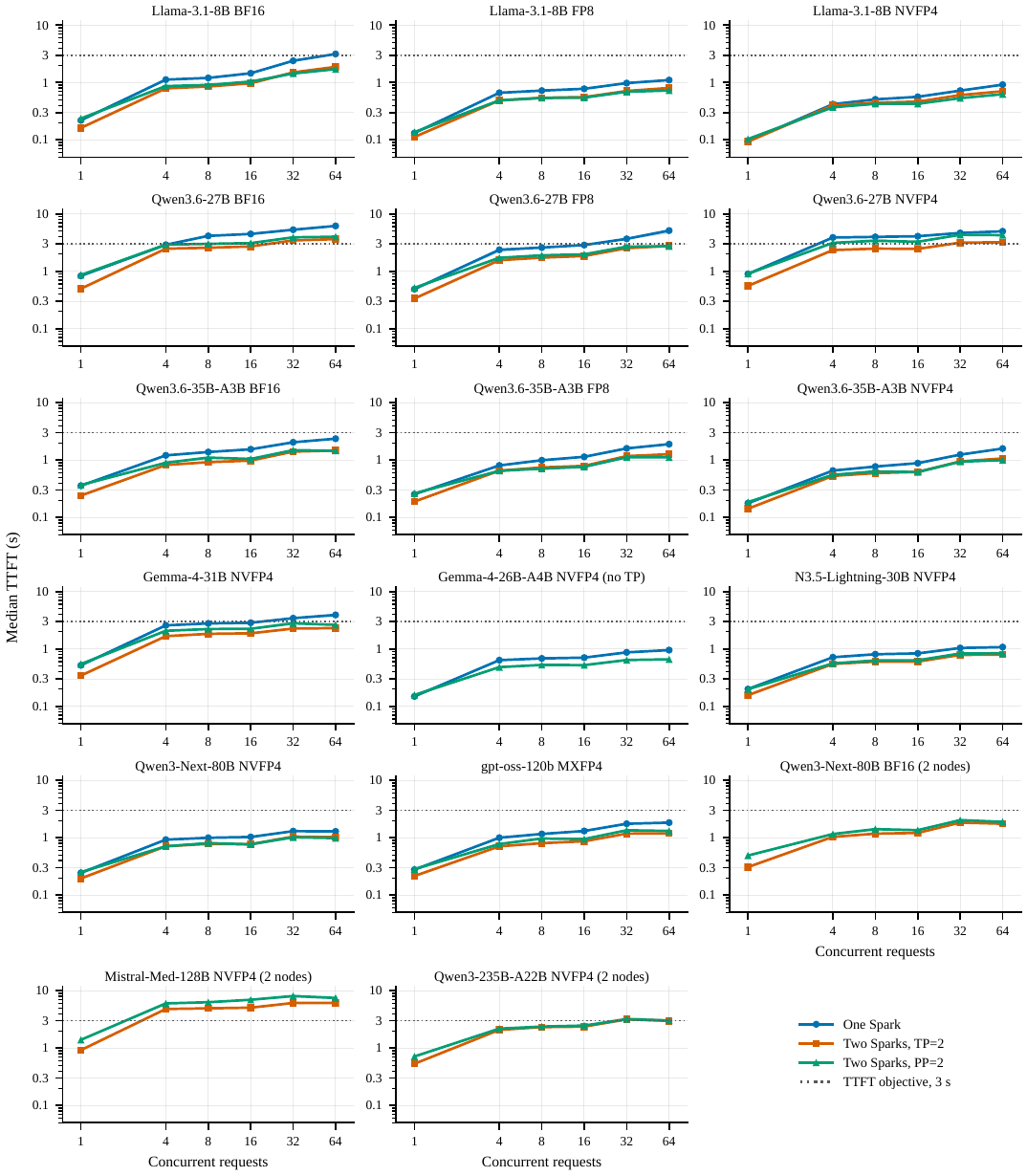}
\caption{Median time to first token (TTFT) versus concurrency on a shared logarithmic scale. The horizontal reference line indicates the \dgxTTFTSLOs\,s TTFT objective.}
\label{fig:ttft}
\end{figure*}

\begin{figure*}[p]
\centering
\includegraphics[width=\textwidth]{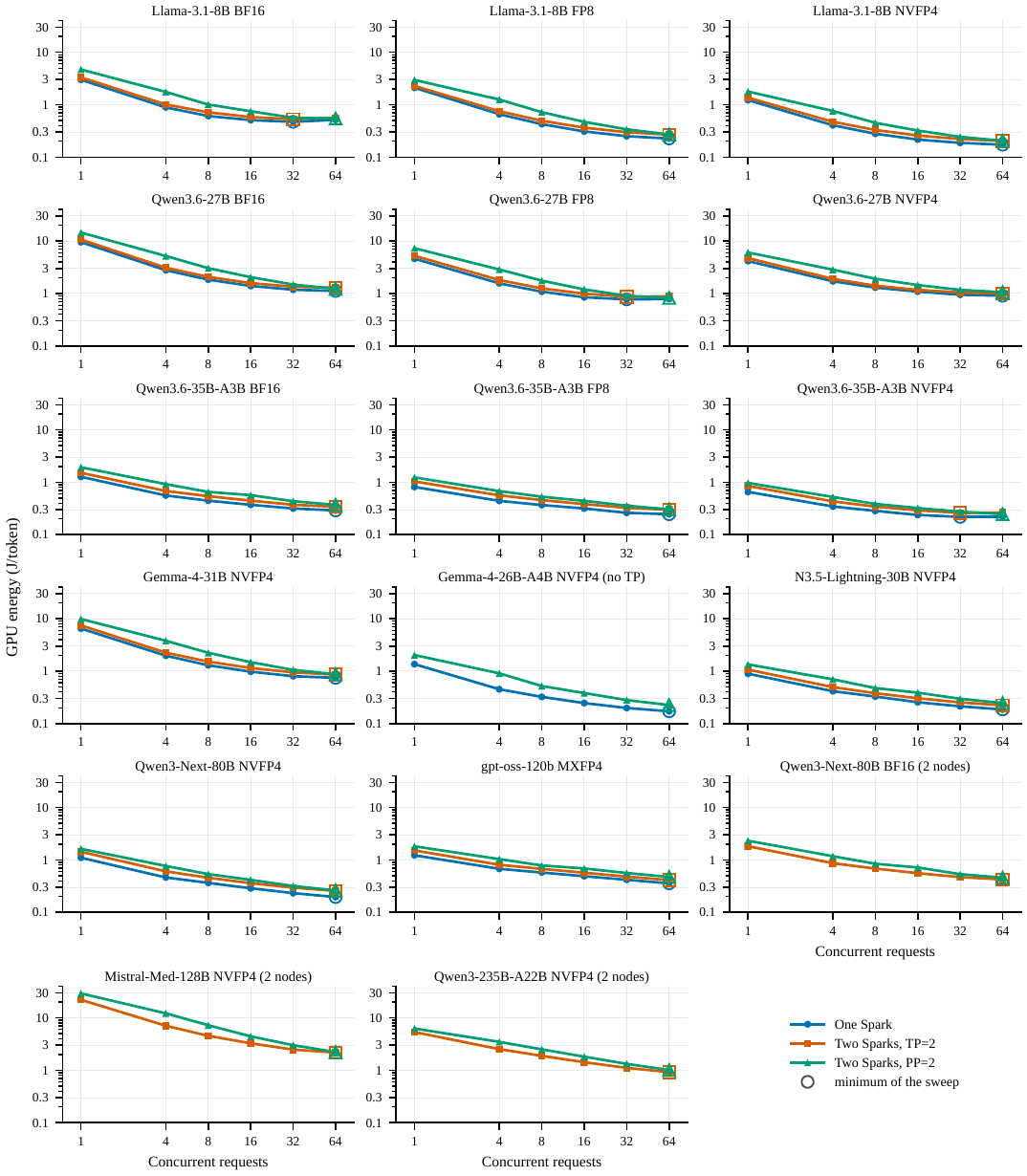}
\caption{GPU energy per generated token versus concurrency on a shared logarithmic scale. Two-Spark measurements include both GPUs. Open markers indicate the minimum energy per token for each deployment curve.}
\label{fig:energygrid}
\end{figure*}

Figs.~\ref{fig:throughput}--\ref{fig:energygrid} present the complete per-checkpoint results supporting Section~\ref{sec:results}, using a consistent layout across metrics. The first three rows show the BF16, FP8 and NVFP4 precision ladders, with checkpoints requiring two Sparks placed last. Tables~\ref{tab:energy} and \ref{tab:crosscampaign} provide additional numerical results for the energy and interconnect analyses.

Two observations warrant attention. First, median TTFT decreases from $c{=}32$ to $c{=}64$ on \dgxTTFTNonMonotoneCurves\ of \dgxNumDualCurves\ two-Spark curves (Fig.~\ref{fig:ttft}), indicating non-monotonic latency behavior under increasing load. Second, at $c{=}1$, the 8{,}192-token bucket contains fewer than ten requests per measurement window on \dgxLongBucketCurvesUnderTen\ of \dgxNumCurves\ curves. These limited samples contribute to the wider replicate ranges observed for long-prompt measurements in Fig.~\ref{fig:prefill}.

\begin{table*}
\caption{Minimum GPU energy per generated token across tested concurrency levels, relative energy consumption compared with one Spark and mean GPU power at $c{=}64$. Power measurements are summed across both nodes for two-Spark deployments.}
\label{tab:energy}
\centering
\footnotesize
\setlength{\tabcolsep}{4pt}
\begin{tabular}{@{}l r r r r r r r r@{}}
\toprule
 & \multicolumn{3}{c}{Min.\ J/token} & \multicolumn{2}{c}{Ratio} & \multicolumn{3}{c}{\shortstack{GPU power\\at $c{=}64$ (W)}} \\
\cmidrule(lr){2-4}\cmidrule(lr){5-6}\cmidrule(lr){7-9}
Checkpoint & $1\times$ & TP & PP & TP/$1\times$ & PP/$1\times$ & $1\times$ & TP & PP \\
\midrule
Llama-3.1-8B BF16 & 0.47 & 0.53 & 0.55 & 1.12 & 1.16 & 63 & 115 & 111 \\
Llama-3.1-8B FP8 & 0.23 & 0.27 & 0.27 & 1.18 & 1.20 & 73 & 125 & 119 \\
Llama-3.1-8B NVFP4 & 0.17 & 0.20 & 0.21 & 1.18 & 1.19 & 67 & 107 & 107 \\
Qwen3.6-27B BF16 & 1.11 & 1.29 & 1.23 & 1.15 & 1.10 & 71 & 136 & 114 \\
Qwen3.6-27B FP8 & 0.77 & 0.87 & 0.83 & 1.12 & 1.07 & 62 & 121 & 111 \\
Qwen3.6-27B NVFP4 & 0.91 & 1.00 & 1.06 & 1.10 & 1.17 & 73 & 132 & 114 \\
Gemma-4-31B NVFP4 & 0.74 & 0.86 & 0.87 & 1.15 & 1.17 & 70 & 136 & 115 \\
Mistral-Medium-3.5-128B NVFP4 & -- & 2.18 & 2.22 & -- & -- & -- & 142 & 115 \\
Qwen3.6-35B-A3B BF16 & 0.29 & 0.34 & 0.37 & 1.18 & 1.28 & 45 & 87 & 79 \\
Qwen3.6-35B-A3B FP8 & 0.25 & 0.30 & 0.31 & 1.22 & 1.24 & 48 & 90 & 86 \\
Qwen3.6-35B-A3B NVFP4 & 0.22 & 0.26 & 0.25 & 1.19 & 1.13 & 53 & 97 & 92 \\
Gemma-4-26B-A4B NVFP4 & 0.17 & -- & 0.23 & -- & 1.32 & 65 & -- & 107 \\
Nemotron-3.5-Lightning-30B-A3B NVFP4 & 0.19 & 0.22 & 0.25 & 1.20 & 1.32 & 64 & 110 & 101 \\
Qwen3-Next-80B-A3B NVFP4 & 0.20 & 0.26 & 0.26 & 1.31 & 1.33 & 56 & 99 & 95 \\
Qwen3-Next-80B-A3B BF16 & -- & 0.42 & 0.46 & -- & -- & -- & 91 & 81 \\
gpt-oss-120b MXFP4 & 0.36 & 0.41 & 0.47 & 1.15 & 1.32 & 67 & 122 & 103 \\
Qwen3-235B-A22B NVFP4 & -- & 0.92 & 1.01 & -- & -- & -- & 115 & 108 \\
\bottomrule
\end{tabular}

\end{table*}

\begin{table}
\caption{Performance comparison of the Llama-3.1-8B precision variants between the July 2026 TCP campaign and the present RDMA campaign (Fig.~\ref{fig:crosscampaign}). Ratios represent RDMA values divided by TCP values.}
\label{tab:crosscampaign}
\centering
\footnotesize
\setlength{\tabcolsep}{3pt}
\begin{tabular}{@{}l l r r r r r r@{}}
\toprule
 & & \multicolumn{3}{c}{Peak tok/s ($c\le64$)} & \multicolumn{3}{c}{ITL at $c{=}1$ (ms)} \\
\cmidrule(lr){3-5}\cmidrule(lr){6-8}
Checkpoint & Deployment & TCP & RDMA & ratio & TCP & RDMA & ratio \\
\midrule
Llama-3.1-8B BF16 & 1$\times$ & 130 & 131 & 1.01 & 72 & 72 & 1.01 \\
 & TP=2 & 130 & 218 & 1.68 & 48 & 38 & 0.78 \\
 & PP=2 & 197 & 204 & 1.04 & 75 & 73 & 0.97 \\
Llama-3.1-8B FP8 & 1$\times$ & 323 & 321 & 0.99 & 39 & 41 & 1.04 \\
 & TP=2 & 197 & 466 & 2.37 & 29 & 22 & 0.77 \\
 & PP=2 & 410 & 439 & 1.07 & 42 & 42 & 0.99 \\
Llama-3.1-8B NVFP4 & 1$\times$ & 390 & 390 & 1.00 & 25 & 26 & 1.01 \\
 & TP=2 & 205 & 529 & 2.58 & 25 & 15 & 0.60 \\
 & PP=2 & 487 & 522 & 1.07 & 29 & 26 & 0.90 \\
\bottomrule
\end{tabular}

\end{table}

\section{Configuration}
\label{app:config}

\subsection{Engine and cluster}

Table~\ref{tab:config} summarizes the serving configuration shared across launches. vLLM used its V1 engine with \texttt{torch.compile} and the Inductor backend, CUDA graphs in \texttt{FULL\_AND\_PIECEWISE} mode and chunked prefill. KV-cache precision was selected automatically: \texttt{fp8\_e4m3} when specified by the checkpoint and the model data type otherwise. Both \texttt{-{}-max-num-seqs} and \texttt{-{}-max-num-batched-tokens} retained their engine defaults. Table~\ref{tab:kernels} reports the quantization formats and kernel backends recorded during engine initialization.

\textbf{Quantization kernels.}
The ModelOpt configurations of \dgxWeightOnlyCountWord\ checkpoints declare weight-only NVFP4 layers (\texttt{W4A16\_NVFP4}). For \dgxMarlinNVCountWord\ checkpoints (\dgxMarlinNVNames), vLLM logged Marlin as the only selected 4-bit kernel, including expert layers in MoE models. This kernel stores weights in 4-bit format while using 16-bit activations, rather than native FP4 computation. In contrast, \dgxNativeNVCountWord\ other NVFP4 checkpoints used the native FlashInfer CUTLASS implementation. One checkpoint (\dgxWeightOnlyNativeNames) declared weight-only layers but logged only the native kernel. These observations are based on kernel-selection logs rather than direct kernel profiling.

Marlin also emitted a warning that native FP4 computation was unsupported and that weight-only execution could reduce compute-intensive performance. This warning is generated by the kernel implementation and does not establish that the Blackwell GPU lacks native FP4 support.

\textbf{Distributed execution and transport.}
Dual-node deployments used Ray's distributed executor, with one container per node. NCCL debug logs identified four RoCEv2 interfaces:

\begin{center}
\footnotesize
\texttt{NET/IB : Using [0]rocep1s0f0:1/RoCE}\\
\texttt{[1]rocep1s0f1:1/RoCE [2]roceP2p1s0f0:1/RoCE}\\
\texttt{[3]roceP2p1s0f1:1/RoCE}
\end{center}

Per-channel logs reported \texttt{NET/IB/<n>} without the \texttt{GDRDMA} suffix. The transport audit verified \dgxAuditIBChannels\ channel records across \dgxAuditLaunches\ dual-node launches. NCCL microbenchmark logs further confirmed that communication used all four RoCEv2 interfaces across both physical cables (Section~\ref{sec:transport}).

Before each launch, the page cache was cleared on both nodes. Checkpoints were downloaded to the head node, copied to the worker, benchmarked and removed before proceeding to the next checkpoint.

\begin{table*}[!tp]
\caption{Serving configuration shared across launches, obtained from serving-policy records, cluster topology and benchmark launch scripts.}
\label{tab:config}
\centering
\footnotesize
\begin{tabular}{@{}>{\raggedright\arraybackslash}p{0.17\textwidth} >{\raggedright\arraybackslash}p{0.76\textwidth}@{}}
\toprule
Item & Value \\
\midrule
Container image &
\texttt{nvcr.io/nvidia/vllm} (NGC release 26.09) on both nodes, digest\newline
\texttt{sha256:557747337846eecf34486866f3d992383c6f8af869fdb0e365b5093fcaee02c2} \\
Engine &
vLLM \dgxVLLMVersion\ (build \texttt{0.29.0+5013de39.nv26.9}), PyTorch 2.14.0a0 (build \texttt{nv26.9}), CUDA runtime 13.4, Python 3.12.3 \\
Kernel libraries &
FlashInfer 0.6.18.post1, flash-attn 2.8.3.post1 \\
Cluster software &
Ray 2.56.1, NCCL 2.31.2 \\
GPU and driver &
NVIDIA GB10, driver 580.178.04, compute capability 12.1, application clock 2{,}418\,MHz on both nodes \\
Head operating system &
Ubuntu 24.04.5 LTS, kernel \texttt{7.0.0-1019-nvidia} (aarch64); the container runs Ubuntu 24.04.4 LTS \\
Engine flags, all launches &
\texttt{-{}-dtype auto}, \texttt{-{}-max-model-len 8704}, \texttt{-{}-gpu-memory-utilization 0.8}, \texttt{-{}-generation-config vllm}, \texttt{-{}-language-model-only}, \texttt{-{}-no-enable-log-requests}, \texttt{-{}-no-enable-prefix-caching}, \texttt{-{}-no-enable-flashinfer-autotune}, and the pinned \texttt{-{}-revision} and \texttt{-{}-tokenizer-revision} of the checkpoint \\
Two-Spark flags &
\tp: \texttt{-{}-tensor-parallel-size 2 -{}-pipeline-parallel-size 1};\newline
\pp: \texttt{-{}-tensor-parallel-size 1 -{}-pipeline-parallel-size 2};\newline
both with \texttt{-{}-distributed-executor-backend ray} \\
Per-checkpoint flag &
\texttt{-{}-trust-remote-code} for the two Qwen3-Next checkpoints \\
Network &
four RoCEv2 rails per node, two on each physical ConnectX-7 port (\texttt{enp1s0f0np0} and \texttt{enP2p1s0f0np0} on port 0, \texttt{enp1s0f1np1} and \texttt{enP2p1s0f1np1} on port 1); one IPv4 /24 subnet per rail; MTU 4096 \\
NCCL environment &
\texttt{NCCL\_IB\_DISABLE=0}; \texttt{NCCL\_SOCKET\_IFNAME} and \texttt{GLOO\_SOCKET\_IFNAME} set to the four rails; \texttt{NCCL\_SOCKET\_NTHREADS=1}; \texttt{NCCL\_NSOCKS\_PERTHREAD=1}; \texttt{NCCL\_DEBUG=INFO}; \texttt{NCCL\_DEBUG\_SUBSYS=INIT,NET}; \texttt{NCCL\_IB\_HCA}, \texttt{NCCL\_IB\_GID\_INDEX}, and \texttt{NCCL\_NET\_PLUGIN} empty \\
Other environment &
\texttt{TORCHINDUCTOR\_AUTOGRAD\_CACHE=0} \\
Workload &
(input, output) token buckets $(128,128)$, $(1024,256)$, $(4096,256)$, $(8192,512)$ with weights $1$, $1$, $1$, $0.5$; closed loop at $c\in\{1,4,8,16,32,64\}$ \\
Window &
warmup of at least 120\,s; measurement of at least \dgxWindowS\,s, extended in 60\,s steps to at least \dgxMinWindowRequests\ started requests (3{,}600\,s cap); cooldown of at least 30\,s \\
Harness &
methodology \texttt{steady-exact-v4}, source commit \texttt{9c14fab} \\
\bottomrule
\end{tabular}

\end{table*}

\begin{table*}[!tp]
\caption{Checkpoint quantization formats and kernel backends reported during vLLM initialization. ModelOpt formats include NVFP4, FP8 and weight-only NVFP4 (W4A16), with $+$ indicating mixed formats. Other checkpoints use block FP8 or MXFP4. A dash indicates that no corresponding kernel selection was logged.}
\label{tab:kernels}
\centering
\footnotesize
\setlength{\tabcolsep}{4pt}
\begin{tabular}{@{}l l l l l l@{}}
\toprule
Checkpoint & Quantization & 4-bit linear kernel & FP8 linear kernel & MoE backend & Attention \\
\midrule
Llama-3.1-8B BF16 & none & -- & -- & -- & FlashAttention \\
Llama-3.1-8B FP8 & FP8 & -- & FlashInfer scaled MM & -- & FlashInfer \\
Llama-3.1-8B NVFP4 & NVFP4 & FlashInfer CUTLASS & -- & -- & FlashInfer \\
Qwen3.6-27B BF16 & none & -- & -- & -- & FlashAttention \\
Qwen3.6-27B FP8 & block FP8 & -- & CUTLASS block-scaled & -- & FlashAttention \\
Qwen3.6-27B NVFP4 & NVFP4+W4A16+FP8 & Marlin, weight-only & FlashInfer scaled MM & -- & FlashInfer \\
Gemma-4-31B NVFP4 & NVFP4 & FlashInfer CUTLASS & -- & -- & -- \\
Mistral-Medium-3.5-128B NVFP4 & NVFP4+W4A16+FP8 & FlashInfer CUTLASS & FlashInfer scaled MM & -- & FlashInfer \\
Qwen3.6-35B-A3B BF16 & none & -- & -- & -- & FlashAttention \\
Qwen3.6-35B-A3B FP8 & block FP8 & -- & CUTLASS block-scaled & -- & FlashAttention \\
Qwen3.6-35B-A3B NVFP4 & NVFP4+W4A16+FP8 & Marlin, weight-only & FlashInfer scaled MM & Marlin & FlashAttention \\
Gemma-4-26B-A4B NVFP4 & NVFP4 & -- & -- & FlashInfer CUTLASS & -- \\
Nemotron-3.5-Lightning-30B-A3B NVFP4 & NVFP4+W4A16+FP8 & Marlin, weight-only & FlashInfer scaled MM & Marlin & FlashInfer \\
Qwen3-Next-80B-A3B NVFP4 & NVFP4 & FlashInfer CUTLASS & -- & FlashInfer CUTLASS & FlashInfer \\
Qwen3-Next-80B-A3B BF16 & none & -- & -- & -- & FlashAttention \\
gpt-oss-120b MXFP4 & MXFP4 & -- & -- & Marlin & FlashInfer \\
Qwen3-235B-A22B NVFP4 & NVFP4 & FlashInfer CUTLASS & -- & FlashInfer CUTLASS & FlashInfer \\
\bottomrule
\end{tabular}

\end{table*}

\subsection{Sampling and windows}

GPU power and temperature were sampled using \texttt{nvidia-smi} at \dgxPowerRateHead\ samples/s on the head node and \dgxPowerRateWorker\ samples/s on the worker. An in-container Ray actor handled worker sampling at a lower frequency to limit overhead. Maximum sampling gaps were \dgxPowerGapHeadMax\,s and \dgxPowerGapWorkerMax\,s, respectively. GPU energy was calculated by interpolating power at measurement-window boundaries and integrating using the trapezoidal rule.

Maximum observed GPU temperatures were \dgxTempHeadMax\,$^\circ$C on the head and \dgxTempWorkerMax\,$^\circ$C on the worker. The highest 99th-percentile end-to-end request latency across measurement windows was \dgxMaxEndToEnd\,s (\dgxMaxEndToEndModel, \dgxMaxEndToEndConfig, $c{=}\dgxMaxEndToEndLoad$) and the longest warmup lasted \dgxMaxWarmupS\,s.

\subsection{Provenance}

Each measurement window records the engine configuration, pinned model and tokenizer revisions, container digest, driver and CUDA versions, GPU clocks, client Python environment and serving-policy fingerprint. SHA-256 digests link each summary to its raw request, power and metrics logs. The harness source is also fingerprinted and the campaign index rejects records with mismatched fingerprints.

A campaign manifest contains SHA-256 digests for \dgxSealedArtifacts\ artifacts. Before generating results, the analysis scripts verify all \dgxSealedChecked\ required index, summary, serving-policy and microbenchmark files against the manifest, terminating if any mismatch is detected.

\section{Artifact}
\label{app:artifact}

The benchmark harness, campaign index, per-window summaries, serving configurations, NCCL microbenchmarks and transport audit are publicly available at \url{https://github.com/kevinhbhimani/dgx-spark-llm-serving-benchmark}. The RDMA campaign is stored under \path{results/RDMA-Betelgeuse-Suite-benchmark-2026-10-30-001}. Raw request-level and GPU power logs are archived separately due to their size and are available from the corresponding author upon request. Reproducing the complete campaign requires exclusive access to two GB10 systems for approximately eleven days.

\clearpage

\end{document}